%% file: main.tex
\documentclass[a4paper,UKenglish,cleveref, autoref, thm-restate]{lipics-v2021}

\input{includes}

\input{defines}

\def\fullversion{1}

\title{Testing Dynamic Environments: Back to Basics}

\author{Yonatan Nakar}{Tel Aviv University, Israel}{yonatannakar@mail.tau.ac.il}{}{}

\author{Dana Ron}{Tel Aviv University, Israel}{danaron@tau.ac.il}{}{}

\authorrunning{Y. Nakar and D. Ron}

\Copyright{Y. Nakar and D. Ron}



\category{} 

\relatedversion{} 

\nolinenumbers 

\EventEditors{John Q. Open and Joan R. Access}
\EventNoEds{2}
\EventLongTitle{42nd Conference on Very Important Topics (CVIT 2016)}
\EventShortTitle{CVIT 2016}
\EventAcronym{CVIT}
\EventYear{2016}
\EventDate{December 24--27, 2016}
\EventLocation{Little Whinging, United Kingdom}
\EventLogo{}
\SeriesVolume{42}
\ArticleNo{23}

\begin{document}

\maketitle

\begin{abstract}
We continue the line of work initiated by Goldreich and Ron ({\em Journal of the ACM, 2017\/}) on testing dynamic environments and propose to pursue a systematic study of the complexity of testing basic dynamic environments and local rules.
As a first step, in this work we focus on dynamic environments that correspond to elementary cellular automata that evolve according to threshold rules.

Our main result is the identification of a set of conditions on local rules, and a meta-algorithm that tests evolution according to local rules that satisfy the conditions.
The meta-algorithm has query complexity $ \poly(1/\eps) $, is non-adaptive and has one-sided error.
We show that all the threshold rules  satisfy the set of conditions, and therefore are $ \poly(1/\eps) $-testable.
We believe that this is a rich area of research and suggest a variety of open problems and natural research directions that may extend and expand our results.
\end{abstract}

\definecolor{bettergreen}{rgb}{0.0, 0.5, 0.0}
\newcommand{\dnew}[1]{{\color{bettergreen}{#1}}}
\newcommand{\dremove}[1]{{\color{gray}{#1}}}

\section{Introduction}

Property testing~\cite{RS,GGR} is the study of algorithms that distinguish between objects that have a given property and those that are far from having the property, by performing a small number of queries to the object.
Goldreich and Ron~\cite{GR-dyn} initiated the study of testing \emph{dynamic environments}, which introduces a temporal aspect to property testing. In this context, the entity being tested changes with time, and is referred to as an \emph{environment}.

Starting from some initial \emph{configuration} (say, a vector or a matrix), the environment is supposed to evolve according to a prespecified \emph{local} rule. The rule is local in the sense that the value associated with each location in the environment at time $ t $ is determined by the values of nearby locations at time $ t-1 $.
The goal of the testing algorithm is then to distinguish between the case that the environment indeed evolves according to the rule, and the case in which the evolution significantly strays from obeying the rule. To this end, the algorithm can query the value of any location of the environment at any of the available time steps, as long as it does not ``go back in time''. Namely, the algorithm cannot choose to query a location at time $ t $ after it has queried some location at time $ t'>t $. We refer to this as the \emph{time-conforming} requirement.
The aim is to design time-conforming algorithms with low query complexity.

Goldreich and Ron~\cite{GR-dyn} investigate the complexity landscape of testing dynamic environments from multiple angles.
From a hardness perspective, they show that there are dynamic environments whose testing requires high query complexity and running time, and that adaptivity and time-conformity are relevant constraints which can significantly impact the query complexity.
However, as we discuss in \Cref{subsec:GR-dyn}, relatively little is known regarding positive results for testing specific rules.

In our quest for understanding which natural families of dynamic environments can be tested efficiently, we propose to first ``go back to the basics'' and study testing in the simplest of dynamic environments.
Namely, in this work we consider environments defined by one-dimensional configurations, which evolve according to local rules that are functions of the current location and its two immediate neighbors.
These dynamic environments, originally introduced by von Neumann~\cite{vonNeumann_automata}, have been extensively studied under the name of \emph{Elementary Cellular Automata}~\cite{wolfram_book} (see definition in \Cref{subsec:intro-defs}).
While these environments can be described in simple terms, they are nevertheless able to capture complex behavior.\footnote{Some rules are even Turing complete~\cite{rule110_universality}.}
Cellular automata have played a role in  various research fields and applications. Examples include modeling physical~\cite{ChopardDroz} and chemical~\cite{model_chemical_systems} systems, VLSI design~\cite{VLSI-ref}, music generation~\cite{music}, analyzing
plant population dynamics~\cite{model_plant_population}, forest fire spread~\cite{model_forest_fire_spread}, city traffic~\cite{model_city_traffic}, urban sprawl~\cite{model_urban_sprawl}, and more.

As we discuss in \Cref{subsec:GR-dyn}, there are several hardness results (both regarding the query complexity and the running time) for testing dynamic environments that correspond to one-dimensional cellular automata (over non-binary alphabets)~\cite{GR-dyn}. Hence, in order to obtain efficient algorithms, it is necessary to restrict the rules considered.
In the current work, our main focus is on perhaps the most basic and natural rules, defined by threshold functions. Such functions have received much attention within the study of propagation of information/influence in networks (see, e.g., the review paper of Peleg~\cite{Peleg-review}, and the recent Ph.D. thesis of Zehmakan~\cite{ZehmakanPhD} and references within).

Our testers are based on a general meta-algorithm which works for rules that satisfy a set of conditions that we define.
In essence, the conditions capture a certain type of behavior leading to ultimate convergence. This behavior induces a global structure on the environment which we exploit in our meta-algorithm.

We hope this work can serve as a basis for further extensions and generalizations, some of which we discuss shortly in \Cref{subsec:future}.

\subsection{Testing basic evolution rules}\label{subsec:intro-defs}

We now formally define the problems we study. We use $ \nums{m} $ to denote the set $ \set{0,...,m-1} $.
For two integers $n$ and $m$, let $\E : \nums{m}\times \cycnums{n} \to \bitset$ denote the evolving environment, and for any $t \in \nums{m}$ let $\E_t :\cycnums{n}\to\bitset$ (the environment at time $t$) be defined by $\E_t(i) = \E(t,i)$.
In general, we refer to a function $\sigma : \cycnums{n} \to \bitset$ as a \emph{configuration}.
When convenient, we may view $\sigma$ as a (cyclic) binary string of length $n$.

For a function (evolution rule) $\rho:\bitset^3 \to \bitset$,  we say that $\E$ \emph{evolves according to $\rho$}, if for every $i \in \cycnums{n}$ and $t>0$, we have that $\E_t(i) = \rho(\E_{t-1}(i-1),\E_{t-1}(i),\E_{t-1}(i+1))$, where all operations are modulo $n$.
We use $\calE^\rho_{m,n}$ to denote the set of environments ${\E : \nums{m}\times \cycnums{n} \to \bitset}$ that evolve according to $\rho$.
As in \cite{GR-dyn}, we employ the standard 
notion of distance used in property testing and say that $\E : \nums{m}\times \cycnums{n} \to \bitset$ is \emph{$\eps$-far from evolving according to $\rho$} ($\eps$-far from $\calE^\rho_{m,n}$)
if $|\{(t,i): \E(t,i) \neq \E'(t,i)\}| > \eps m n$ for every $\E'\in \calE^\rho_{m,n}$.\footnote{In the context of dynamic environments, this notion of distance can be interpreted as capturing ``measurement errors'' due to some noise process.
	Namely, it can be viewed as allowing the testing algorithm to accept not only ``perfect'' environments, but also environments that correspond to a correct evolution with a bounded fraction of corruptions.
	Also note that being $\eps$-far from evolving according to $\rho$ does not simply translate to there being an $\eps$-fraction of pairs $(t,i)$ for which $\E_t(i) \neq \rho(\E_{t-1}(i-1),\E_{t-1}(i),\E_{t-1}(i+1))$ (which would be trivial to test).
}

Given $n$, $m$, and a distance parameter $\eps \in (0,1)$, a testing algorithm for evolution according to a rule $\rho$ should distinguish with constant success probability between the case that an environment $\E$ belongs to $\calE^\rho_{m,n}$ and the case that it is $\eps$-far from $\calE^\rho_{m,n}$. To this end, the algorithm is given query access to $\E$, where a query on a pair $(t,i)$ cannot follow any query on $(t',i')$ for $t' > t$. We are interested in bounding both the total number of queries performed by the algorithm (as a function of $\eps$, and possibly $m$ and $n$) and the maximum number of queries it performs at any time step (which we refer to as its \emph{temporal query complexity}).

\subsection{Our results}\label{subsec:results}

We identify several conditions on local rules (which are formally stated in~\Cref{subsec:conditions}),
such that if a rule $\rho$ satisfies these conditions, then evolution according to $\rho$ can be tested with query complexity $\poly(1/\eps)$ with one-sided error.
Our testers have the advantage that they are non-adaptive, and therefore, in particular, time-conforming.

\begin{theorem}\label{thm:cond-meta}
	Let $\Psi$ be the set of conditions specified in \Cref{subsec:conditions}. For every rule $\rho$ that satisfies the conditions in $\Psi$, it is possible to test evolution according to $\rho$ by performing $O(1/\eps^4)$ queries.
	Furthermore, the testing algorithm is non-adaptive and has one-sided error.
\end{theorem}

To establish \Cref{thm:cond-meta}, we present a \emph{meta-algorithm} for testing evolution and prove its correctness for rules that satisfy the aforementioned conditions (the set $\Psi$). It is a meta-algorithm in the sense that it is based on certain subroutines that are rule-specific (but have a common functionality of detecting violations of evolution according to the tested rule). We provide a high-level discussion of the conditions and the algorithm in \Cref{subsec:high-level}.

\medskip\noindent
Our main application of the meta-algorithm is to the natural family of threshold rules.

\setcounter{definition}{0}
\begin{definition}\label{definition:threshold}
	We say that a rule $ \rho :\bitset^3 \to \bitset $ is a \emph{threshold} rule if there exist a threshold integer
	$ 0 \leq b \leq 3 $ and a bit $\alpha \in \set{0, 1} $ such that $ \rho(\beta_1,\beta_2,\beta_3) = \alpha $ if and only if $ \beta_1+\beta_2+\beta_3 \geq b $.
\end{definition}

We prove:

\setcounter{theorem}{1}
\begin{theorem}\label{thm:threshold-test}
	For each threshold rule $\rho$, evolution according to $\rho$ can be tested with query complexity $O(1/\eps^4)$. Furthermore, the testing algorithm is non-adaptive and has one-sided error.
\end{theorem}

We also show that the conditions hold for two additional (non-threshold) rules, so the applicability of our meta-algorithm is more general (\ifnum\fullversion=1 \Cref{subsec:other-rules}\else for details, see full version of this paper~\cite{fullversion}\fi).
We believe that appropriate (perhaps more complex) variants of our algorithm can be used to test an even larger variety of basic local rules (see \Cref{subsec:future}), where we conjecture that this is true for all rules that ultimately converge.
Interestingly, while the two additional rules are not threshold rules as per \Cref{definition:threshold}, they can be represented as weighted threshold rules (which are a subclass of ultimately converging rules).

\subsection{The high-level ideas behind our results}\label{subsec:high-level}
In this high-level discussion, we assume for simplicity that $m\geq n$ (the case $m<n$ can be essentially reduced to this case).

\subsubsection{Convergence, final/non-final locations and prediction functions}

To give an intuition on the convergence behavior that our conditions capture, it is useful to first discuss the notion of \emph{ultimate convergence}.
A rule $\rho$ ultimately converges if, for any initial configuration $\E_0$, an environment evolving according to $\rho$ converges after a bounded number of steps to either a single final configuration or to a constant number of configurations between which it alternates. For example,  consider the majority rule (threshold $2$). Unless the initial configuration is $(01)^{n/2}$, the environment ultimately converges to some configuration that consists of blocks of $0$s and $1$s of size at least 2 each (and if it is $(01)^{n/2}$, then it alternates between $(01)^{n/2}$ and $(10)^{n/2}$).

Once an environment converges, testing is straightforward since we can easily predict the values of locations in future time steps and then verify that indeed they hold the predicted values (or else we reject).
The issue, however, is that convergence is not ensured to be reached after a small number of time steps.\footnote{In fact, there are initial configurations that require $\Omega(n)$ steps before they ultimately converge.}
In other words, knowing that a rule ultimately converges cannot be exploited directly.
Hence, the challenge is to identify and formalize conditions that allow for ``pre-convergence prediction''. Namely, conditions that imply the ability to predict future values of locations based on the current values of these and other locations (before convergence is reached).

In this context, our conditions try to formalize the idea that rules exhibit a certain \emph{local} convergence, which ``expands'' with time.
The first ingredient of our approach is the observation that, in the case of the majority rule, if at any time step $t$, $\E_t(i) \in \{ \E_t(i-1), \E_t(i+1)\}$ , then $\E_{t'}(i) = \E_t(i)$ for any $t' > t$ (operations  are modulo $n$). We say in such a case that location $i$ is \emph{final} at time $t$ (in $\E$). Otherwise it is \emph{non-final}. Crucially for us, whether a location $i$ is final or not  at a certain time step depends solely on its local neighborhood at that time (and can hence be verified with a constant number of queries).

An important property of a location being final at time $t$ (in addition to converging to their final value, up to alternations), is the aforementioned expansion (or ``transmission of finality''). Namely, a location $i$ that is non-final at time $t$ becomes final at time $t+1$ if either $i-1$ or $i+1$ is final at time $t$ (possibly both). Furthermore, it cannot become final if both its neighbors are non-final. Another related property of final locations is that (under certain circumstances), they can be used to predict the values of locations that become final in the future, based on a (rule-specific) \emph{prediction function}. A similar statement holds for non-final locations (though the circumstances are different).

\subsubsection{The meta-algorithm: the grid and violating pairs}
Based on these properties (which are formalized in the conditions we introduce), our (meta) algorithm works in two stages. In the first stage, it queries the environment at time $t_1 = \Theta(\eps m)$ on $O(1/\eps^2)$ equally spaced locations, which we refer to as \emph{the grid locations}, and their local neighborhoods. This allows the algorithm to determine which of the grid locations are final at time $t_1$ and which are non-final. If the answers it gets are not consistent with any environment that evolves according to $\rho$ (in which case we say that the grid is \emph{not feasible}), then it rejects.

In its second stage, the algorithm uniformly samples $O(1/\eps)$ random time-location pairs $(t,i)$ and queries $\E_t$ on $i$ and its local neighborhood. It then checks whether the answers are consistent with the answers to queries in the first stage (on the grid locations and their neighborhoods) or constitute a \emph{violation}. The definition of consistency/violation is based on the aforementioned prediction functions of the tested rule.

One may have hoped that such a consistency check is sufficient, in the sense that all (or almost all) pairs $(t,i)$ can be predicted based on the answers to the queried grid locations. Unfortunately, this is not the case. There are (possibly many) pairs $(t,i)$ whose $0/1$ values are not determined given the first-stage answers. However, we show that such pairs are constrained in a different way (in environments that evolve according to $\rho$): their location must have become final by time $t_2 = t_1+\Delta$, where $\Delta$ is the distance between grid location. Hence, for each selected pair $(t,i)$, the algorithm also queries $\E_{t_2}$ on location $i$ (and its neighborhood) and  checks consistency with the queried locations at time $t_2$.

\subsubsection{On the analysis of the algorithm and ``backward prediction''}
To show that the algorithm always accepts environments that evolve according to the tested rule $\rho$, we prove that our definition of violation is such that there are no violations in such environments (assuming $\rho$ satisfies the aforementioned conditions).
The more involved part of the analysis is proving that if the environment $\E$ is $\eps$-far from evolving according to $\rho$, then the algorithm will detect this with probability at least $2/3$. To this end, we prove the contrapositive statement. Namely, we show that if the algorithm accepts with probability at least $2/3$, then there exists an environment that evolves according to $\rho$ and is $\eps$-close to $\E$. This is done by showing that we can construct an initial configuration $\E'_0$, such that if we let it evolve according to $\rho$, resulting in an environment $\E' \in \calE^\rho_{m,n}$, then we can upper bound the number of pairs $(t,i)$ such that $\E_t(i) \neq \E'_t(i)$ by $\eps m n$.

Here we build on a useful property of the prediction functions, by which they allow us a certain ``prediction back in time''. Namely (for $t_1$ and $t_2$ as mentioned above),  we use the queried grid locations at time $t_1$ as well as some locations at time $t_2$ (which have not been queried) to determine values of locations at the earlier time $0$ in $\E'$. We prove that this can be done in a way that ensures that $\E'$ agrees with $\E$ on all pairs $(t,i)$ that are not violating.

\subsection{A short overview of the results in~{\cite{GR-dyn}}}\label{subsec:GR-dyn}

As stated earlier, the study of testing dynamic environments was initiated by Goldreich and Ron~\cite{GR-dyn},
who present several general results as well as analyze two natural specific rules.
We first provide a short overview of their main general results.

They prove that the query complexity of testing (one-dimensional) rules may have high query complexity.
Specifically, they show that there exists a constant $ c > 0 $ and an evolution rule $ \rho : \Sigma^3 \to \Sigma $ such that any tester of evolution according to $ \rho $ requires $ \Omega(n^c) $ queries.\footnote{Observe that it is possible to test the evolution according to any rule $\rho$ over configurations of size $n$
	by performing $O(n+1/\eps)$ queries ($n$ queries to the initial configuration and $O(1/\eps)$ uniformly selected queries elsewhere). To get sublinear temporal query complexity, a total of $O(n/\eps)$ uniformly selected queries suffice (by applying a simple union bound over all possible initial configurations).}
They also prove that testing dynamic environments may be NP-Hard, provided that the temporal query complexity is ``significantly sublinear'' (where $ f(x) $ is significantly sublinear if $ f(x) < x^{1-\Omega(1)} $).
More precisely, they show that for every constant $ c > 0 $ there exists an evolution rule $ \rho : \Sigma^3 \to \Sigma $ such that no (time-conforming) polynomial-time testing algorithm with temporal query complexity $ n^{1-c} $ can test whether $ n $-sized environments evolve according to $ \rho $
(assuming $\mathcal{NP}\not\subseteq \mathcal{BPP}$).
Their general results also include a theorem concerning the usefulness of adaptivity in testing dynamic environments, a study of the relation between testing and learning dynamic environments, and a result on the power of being non time-conforming.

Goldreich and Ron~\cite{GR-dyn} also provide testers for evolution according to two specific (classes of) rules.
The first is the class of linear rules, which in the binary 1-dimensional case corresponds to the XOR rule in elementary cellular automata. They show that for any $ d \ge 1 $ and any field $ \Sigma $ of prime order, there exists a constant $ \gamma < d $ such that the following holds.
For any linear rule $ \rho : \Sigma^{3^d} \rightarrow \Sigma $ there exists a time-conforming oracle machine of (total) time complexity $ \poly(1/\eps) \cdot n^\gamma $ that tests the consistency of an evolving environment with respect to $ \rho $.
Furthermore, the tester is non-adaptive and has one-sided error.

Their second specific positive result, loosely stated, captures fixed-speed movement of objects in one-dimension such that colliding objects stop forever.
They present a (time-conforming) algorithm of (total) time complexity poly($ 1/\eps $) that tests the consistency of evolving environments with respect to that rule.

\subsection{Future directions }\label{subsec:future}

\subparagraph*{Basic dynamic environments.} A natural question that arises is whether a more nuanced version of the set of conditions formalized in this paper and the meta-algorithm can be defined and proved to work for other rules in the realm of basic dynamic environments.
Indeed, preliminary results suggest that several other rules that ultimately converge exhibit behaviors that ``resemble'' the ones captured by our conditions.
This leads us to the following conjecture.

\bigskip\noindent\textsf{Conjecture.}~\textit{
	If a rule $ \rho $ ultimately converges, then it is poly($ \frac{1}{\epsilon} $)-testable.}
\medskip

While our meta-algorithm does not apply to rules that do not ultimately converge, there are natural rules that fall under this category (the XOR rule for instance).
This raises the question of whether $ \poly(1/\eps) $ testers exist for such rules.
The answer is that there are $ \poly(1/\eps) $-testable rules that do not ultimately converge, but as we'll see, the question should be slightly rephrased.\footnote{We thank one of the anonymous reviewers of this paper for pointing this out.}
To give one example, for the rule $ \rho $ defined as $ \rho(x,y,z)=x $, each configuration is simply a copy of the previous configuration, shifted one location to the right.
That is, while an environment evolving according to this rule does not, technically, ultimately converge, this rule is trivially $ \poly(1/\eps) $-testable.
However, this particular rule (and other rules that are capable of producing such ``shifting behaviors'') also has the property of not being \textit{symmetric} (i.e., it does not hold that $ \rho(x,y,z) = \rho(z, y, x) $ for every $ x,y,z $).
Hence, one way to rephrase the question is restricting it to symmetric rules.

\begin{openproblem}
	Are there any symmetric rules that do not ultimately converge and are $ \poly(1/\eps) $-testable?
\end{openproblem}

Another way to rephrase this question is to define a more general notion of ultimate convergence.
Specifically, we say that a rule $\rho$ \textit{ultimately converges up to a shift} if, for any initial configuration $\E_0$, an environment evolving according to $\rho$ converges after a bounded number of steps to a constant number of configuration \textit{equivalence classes} between which it alternates, where two configurations are \textit{equivalent} if they are equal \textit{up to a shift}.

\begin{openproblem}
	Are there any non-symmetric rules that do not ultimately converge up to a shift and are $ \poly(1/\eps) $-testable?
\end{openproblem}

As mentioned in \Cref{subsec:GR-dyn}, it has been shown in~\cite{GR-dyn} that the XOR rule is sublinearly testable.
However, the query complexity of the tester depends on the size of the environment and is only mildly sublinear (the complexity is $ O(n^{0.8}) $ for an environment of size $ n $).
This raises the question of whether there exists a tester for the XOR rule with significantly lower query complexity (maybe even polylogarithmic).
Another question that can be raised is whether there are other symmetric rules, ones that do not ultimately converge, that can be tested with a sublinear query complexity that depends on the size of the environment.

\begin{openproblem}
	Which symmetric rules that do not ultimately converge can be tested with query complexity that is sublinear in (but strictly grows with) the size of the environment?
\end{openproblem}

\subparagraph*{More general dynamic environments.}
Building on the ideas for testing basic dynamic environments, it may be possible to venture into more general environments.
One such generalization is to consider rules that depend on more than just the three locations constituting the immediate neighborhood.
Other generalizations are to environments and rules over non-binary values, higher dimensions, and environments that evolve on more general graphs.

\subparagraph*{Non-deterministic rules.}
We also suggest considering local rules that are non-deterministic in the sense that given some configuration, the rule allows several configurations to follow.
An example of one such rule, which can be thought of as a relaxation of the $ \orr $ rule, is the rule in which each value is restricted to be monotonically non-decreasing with respect to the previous values at the location's neighborhood.

\ifnum\fullversion=0
\subsection*{Missing details}
Due to space constraints, not all details appear in this extended abstract, and can be found in the full version of this paper~\cite{fullversion}.
\fi

\section{Preliminaries}\label{sec:prel}
\setcounter{definition}{1}

In addition to the basic definitions provided in \Cref{subsec:intro-defs} regarding testing dynamic environments, here we introduce several more definitions and notations.

In all that follows, when performing operations on locations $i \in \cycnums{n}$, these operations are modulo $n$.
For a pair of locations $i,j \in \cycnums{n}$ we use
$[i,j]$ to denote the sequence $i,i+1,\dots,j$ (so that it is possible that $j < i$).

\begin{definition}\label{def:neighborhood}
	For a location $i\in \cycnums{n}$ and an integer $r$, the \textsf{$r$-neighborhood} of $i$, denoted $\Gamma_r(i)$, is the sequence $[i-r,i+r]$. For a set of locations $I \subseteq \cycnums{n}$, we let $\Gamma_r(I)$ denote the set of locations in the union of sequences $[i-r,i+r]$ taken over all $i\in I$.
\end{definition}

\begin{definition}\label{def:state-machine}
	For an integer $n$ and a local rule $\rho$,  let $\M_{\rho}(n)$ denote the (deterministic) \textsf{state machine} that is defined as following. Each state of $\M_{\rho}(n)$ corresponds to a different configuration $\sigma : \cycnums{n}  \to \bitset$. 
	If a state corresponds to a configuration $\sigma$, then it has a single transition going to the state corresponding to
	the configuration that results from applying $\rho$ to $\sigma$.

	The \textsf{period} of $\M_\rho(n)$, denoted $p_\rho(n)$, is the longest size of a (directed) cycle in $\M_\rho(n)$. If
	there exists a constant $p$ such that $p_\rho(n) = p$ ($p_\rho(n) \leq p$) for every sufficiently large $n$, then we say that $\rho$ \textsf{has period} (at most) $p$, and that $\rho$ \textsf{ultimately converges}.
\end{definition}

\noindent
Observe that for every $\M_\rho(n)$, each strongly connected component in $\M_\rho(n)$ is either a single state with no edges in the component  or a cycle (where in particular, the cycle may be a self-loop).
For example, if $\rho$ is the OR function, then it has period $1$ (as it contains only two cycles: one is a self-loop for the state corresponding to the configuration $1^n$ and the other is a self loop for the state corresponding to the configuration $0^n$). On the other hand, there are rules, such as XOR, for which $p_\rho(n) = \Omega(n)$.

\begin{definition}\label{def:dist}
	For two locations $i,i' \in \cycnums{n}$, we let
	$\ddist(i,i') = i'-i$ denote the directed distance from $i$ to $i$, and let
	$\dist(i,i') = \min\{\ddist(i,i'),\ddist(i',i)\}$ denote the (undirected) distance.
\end{definition}

\noindent
Note that since operations on locations are modulo $n$, we have that
$\ddist(i,i') \leq n-1$, while
$\dist(i,i') \leq n/2$ for all $i,i'\in \cycnums{n}$.

\begin{definition}\label{def:pair}
	For $t\in \nums{m}$ and $i \in \cycnums{n}$, we refer to $(t,i)$ as a \textsf{time-location pair} (or simply \textsf{pair}).
	
	\label{def:descends}
	Given two locations, $ i,i' \in \cycnums{n} $ and two time steps $ t,t' \in \nums{m} $ where $ t > t' $, we say that the pair $ (t,i) $ \textsf{descends} from the pair $ (t',i') $ if $ \dist(i,i') \le t-t' $.
	We say that $(t,i)$ is a \textsf{descendant} of $(t',i')$ and that $(t',i')$ is an \textsf{ancestor} of $(t,i)$.
\end{definition}

\begin{definition}\label{def:pattern}
	For an integer $r$, an \textsf{$r$-pattern} is a string in $\bitset^r$.
\end{definition}

\section{The Conditions}\label{subsec:conditions}

Let $ \rho :\bitset^3 \to \bitset$ be a local rule. We present several conditions, such that if they all hold, then the rule $ \rho $ can be tested with $ \poly(1/\eps) $ queries.
These conditions capture properties of local rules that can be exploited by our (meta) algorithm.

The conditions are defined with respect to a constant (integer) $k$ (which depends on $\rho$, but for the sake of simplicity we suppress the dependence on $\rho$ and use $k$ rather than $k_\rho$), and a partition of all $(2k+1)$-patterns.\footnote{For the local rules we apply our conditions to, $k$ is either $0$ or $1$, but using a variable parameter $k$ will hopefully allow to extend our results more easily.}
The partition is denoted by $(\F_\rho,\bF_\rho)$, where $\F$ stands for \emph{final} and $\bF$ for \emph{non-final}.

We shall say that a pair $(t,i)$ is final (respectively,  non-final) \emph{with respect to $\E$ and $\rho$} if $\E_t(\Gamma_k(i)) \in \F_\rho$ (respectively, $\bF_\rho$).
Roughly speaking, if $(t,i)$ is final (with respect to $\E$ and $\rho$), then location $i$ does not change from time $t$ and onward (or, more generally, $\E_{t'}(i)$ for $t'>t$ can be predicted based on $\E_t(i)$).
Furthermore, if $(t,i)$ is non-final, then $(t+1,i)$ is final if and only if $(t,i-1)$ or $(t-1,i+1)$ is final (so that finality is ``infectious'').

In our statements of the conditions, we make use of the parity function, which we denote by $ \parity:\NN\to \set{0,1} $ (so that $\parity(x)  = 1$ if $x$ is odd and $\parity(x)=0$ if $x$ is even).

Before each of the conditions is stated formally, we give a short, informal description.
It will also be useful to have a running example of a specific rule $\rho$, which is the majority rule.
Namely, $\maj(\beta_1,\beta_2,\beta_3) = 1$ for any three bits $\beta_1,\beta_2,\beta_3$, if and only if
$\beta_1+\beta_2+\beta_3 \geq 2$. For the majority rule, $k=1$, and $\F_{\maj} =\{111,110,011,000,001,100\}$ (so that $\bF_{\maj} = \{101,010\}$).

The first condition says that if a location is final, then it remains final.

\begin{condition}\label{condition:final}
	Let $\E \in  \calE^\rho_{m,n}$ be an environment that evolves according to $\rho$.
	For any time step $t \in \nums{m-1}$ and location $i\in \cycnums{n}$, if $\E_t(\Gamma_k(i))\in \F_\rho$, then $\E_{t+1}(\Gamma_k(i)) \in \F_\rho$.
\end{condition}
\noindent
Indeed, for the majority rule, if $\E_t(\Gamma_1(i)) = 111$, then $\E_{t+1}(\Gamma_1(i)) = 111 \in \F_{\maj}$,
if $\E_t(\Gamma_1(i)) = 110$, then $\E_{t+1}(\Gamma_1(i)) \in \{110,111\} \subset \F_{\maj}$, and if $\E_t(\Gamma_1(i)) = 110$, then $\E_{t+1}(\Gamma_1(i)) \in \{110,111\} \subset \F_{\maj}$ (analogous statements hold for $\E_t(\Gamma_1(i)) \in \{000,001,100\}$).

\smallskip
The second condition  says that if a location is  non-final, then it can become final in one time step if and only if it has a final neighbor.
\begin{condition}\label{condition:infecting_neighbors}
	Let $\E \in  \calE^\rho_{m,n}$ be an environment that evolves according to $\rho$.
	For any time step $ t \in \nums{m-1} $ and location $ i\in \cycnums{n} $, if $\E_t(\Gamma_k(i)) \in \bF_\rho$, then
	$\E_{t+1}(\Gamma_k(i)) \in \F_\rho$ if and only if $\E_t(\Gamma_k(i-1)) \in \F_\rho$ or
	$\E_t(\Gamma_k(i+1)) \in \F_\rho$ (or both).
\end{condition}
\noindent
For the majority rule, consider the case that $\E_t(\Gamma_1(i)) = 101$ (so that it belongs to $\bF_{\maj})$.
In this case,  $\E_t(\Gamma_1(i-1)) \in \{110,010\}$ and $\E_t(\Gamma_1(i+1))\in \{011,010\}$.
If $\E_t(\Gamma_1(i-1))=110$ (which belongs to $\F_{\maj}$), then $\E_{t+1}(\Gamma_1(i)) \in \{110,111\} \subset \F_{\maj}$,
and the case that $\E_t(\Gamma_1(i+1))=011$ is analogous.
On the other hand, if both $\E_t(\Gamma_1(i-1))=010$ and $\E_t(\Gamma_1(i+1))=010$ (so that they both belong to $\bF_{\maj}$), then
$\E_{t+1}(\Gamma_1(i)) = 010$ (and it belongs to $\bF_{\maj}$ as well).
Note that, if for every location $ i \in \cycnums{n} $, it holds that $\E_0(\Gamma_1(i)) \in \set{010, 101}$ (that is, every location in the initial configuration is non-final), then no location would ever become final throughout the evolution of the rule.
In particular, in this case the environment alternates between $(01)^{n/2}$ and $(10)^{n/2}$, where all the locations are non-final.

\smallskip
The first two conditions intuitively imply that one can determine whether certain locations are final or non-final using particular ``past'' locations that are known to be final or non-final.
The next two conditions capture the idea that the actual \emph{values} at certain locations (and not only whether or not they are final) can also be determined based on past locations.

In particular, the third condition captures how values at locations that are final at a certain time step can be predicted using a function that depends on ``past'' final locations from which they descend (and to which they are closest).

\begin{condition}\label{condition:final_prediction}
	Let $\E \in  \calE^\rho_{m,n}$ be an environment that evolves according to $\rho$.
	There exists a function $ \frhor: \bitset^3 \to \bitset $ for which
	the following holds.
	First, $\frhor$ is the XOR of its first argument and a subset of the other two arguments.
	Second, let $ (t, i) $ and $ (t', i') $ be any two pairs such that $ (t,i) $ descends from $ (t',i') $, $\E_{t}(\Gamma_k(i)) ,\E_{t'}(\Gamma_k(i')) \in \F_\rho$, and for every $ i'' \ne i' $ satisfying $ \dist(i,i'') \le \dist(i,i') $ it holds that $ \E_{t'}(\Gamma_k(i'')) \in \bF_\rho$. Then
	\[ \E_t(i) = \frhor(\E_{t'}(i'), \parity(t-t'), \parity(\dist(i,i'))  \;.\]
\end{condition}
\noindent
For the majority rule, $\fruler{\maj}$ is simply the identity function on its first argument, namely, $\fruler{\maj}(\beta,\cdot,\cdot)=\beta$.

\smallskip
The fourth condition captures how  locations that are non-final at a certain time step can be predicted using a function that depends on ``past'' non-final locations from which they descend (conditioned on there not being any final location among its ancestors in that past time step).

\begin{condition}\label{condition:noninf_prediction}
	Let $\E \in  \calE^\rho_{m,n}$ be an environment that evolves according to $\rho$.
	There exists a function $ \hrhor :\bF_\rho\times \bitset \times \cycnums{n} \to  \bF_\rho$
	for which the following holds. First, $ \hrhor$ is reversible in the sense that
	for each fixed $\tau \in \bF_\rho$, $\beta \in \bitset$ and  $\ell \in \cycnums{n}$, there exists a unique $\tau'$ such that  $\hrhor(\tau',\beta,\ell) = \tau$. Second,  let $ (t, i) $ and $ (t', i') $ be any two pairs such that $ (t,i) $ descends from $ (t',i') $, $\E_{t}(\Gamma_k(i)),\E_{t'}(\Gamma_k(i')) \in  \bF_\rho $, and $ \E_{t'}(\Gamma_k(i'')) \in  \bF_\rho $ for every $ i'' $ such that $ (t,i) $ descends from $ (t',i'') $. Then
	\[ \E_t(\Gamma_k(i)) = \hrhor(\E_{t'}(\Gamma_k(i')), \parity(t-t'), \ddist(i',i)) \;. \]
	
\end{condition}

\noindent
For the majority rule,
$\hruler{\maj}(010,\beta,x)=010$ if $\beta\oplus \parity(x) = 0$ and $\hruler{\maj}(010,\beta,x)=101$ if
$\beta\oplus \parity(x) = 0$. Similarly, $\hruler{\maj}(101,\beta,x)=101$ if $\beta\oplus \parity(x) = 0$
and $\hruler{\maj}(101,\beta,x)=010$ if $\beta\oplus \parity(x) = 1$.

The additional two conditions presented below are a bit more involved than Conditions~\ref{condition:final}--\ref{condition:noninf_prediction}, and perhaps initially less intuitive.
They do not play a role in the definition of the meta algorithm, but are applied in the proof of \Cref{lemma:soundness} (and we recommend that the reader return to them in that context).
In a nutshell, they allow us to show that if our testing algorithm accepts the environment $\E$ with  high constant probability, then there exists an environment $\E'$ that evolves according to $\rho$ and is relatively close to $\E$.
In particular, they aid us in defining the initial configuration $\E'_0$ based on $\E_t$ for some appropriate time step $t$.

\begin{condition}\label{condition:non-localA}
	Let $ \sigma : \cycnums{n} \to \bitset $ be a configuration and let $ [x,y] $ be an interval
	of locations such that $ \sigma(\Gamma_k(x)) \in  \F_\rho $ and $ \sigma(\Gamma_k(y)) \in \F_\rho $.
	There exists a configuration $ \tsigma \in \cycnums{n} $, which differs from $\sigma$ only on locations inside $[x,y]$, for which the following holds:
	For every $i \in [x,y]$ we have that $ \tsigma(\Gamma_k(i)) \in \F_\rho $, and if $\sigma(\Gamma_k(i)) \in \F_\rho$, then $ \tsigma(i)=\sigma(i) $.
	
	This condition also covers the special case in which $y=x$ and we interpret $[x,y]$ as $x,x+1,\dots,x+n$ (with a slight abuse of notation).
\end{condition}

\begin{condition}\label{condition:non-localB}
	Let $ \sigma : \cycnums{n} \to \bitset $ be a configuration and $z \in \cycnums{n}$ such that $ \sigma(\Gamma_k(z)) \in  \bF_\rho $.
	Let $\nu \in \{\tau_{k+1}:\tau \in \F_\rho \}$ and $\gamma,\gamma'\in \bitset$.
	There exists a configuration $ \tsigma : \cycnums{n} \to \bitset $ for which the following hold.
	There is a  location $z' \in [z+1,z+2k+1]$  where $ \tsigma(\Gamma_k(z')) \in \F_\rho $, and  $\frhor(\tsigma(z'),\gamma,\parity(z'-z)\oplus \gamma') = \nu$.
	Furthermore, for every $ i \in [z+1,z'-1]$ it holds that $ \tsigma(\Gamma_k(i)) \in \bF_\rho $, and for every
	$ i \notin [z + k, z'+k] $, $ \tsigma(i)=\sigma(i) $.
	
	A (symmetric) variant of the above should also hold if we replace $z' \in [z+1,z+2k+1]$  by $z' \in [z-2k-1,z-1]$, $ i \in [z+1,z'-1]$ by $i\in [z'+1,z-1]$, and $ i \notin [z + k, z'+k] $ by $i \notin [z' - k, z-k] $.
\end{condition}

\section{The Meta-Algorithm}\label{sec:cond-alg}

In this section, we present a meta-algorithm for testing evolution of local rules that satisfy the sufficient conditions (specified in \Cref{subsec:conditions}).
Here we give an algorithm whose complexity is $\lceil n/m\rceil\cdot \poly(1/\eps)$ and, in \ifnum\fullversion=1\Cref{subsec:m-n}\else the full version of this paper~\cite{fullversion}\fi, we explain how to remove the dependence on $n/m$.

In order to precisely describe our meta-algorithm, we need to first define a particular set of locations that we designate as \textit{the 1-dimensional grid} and the notion of \textit{violating time-location pairs} with respect to the 1-dimensional grid.
The 1-dimensional grid is defined in \Cref{subsec:grid} and the notion of violating pairs is defined in \Cref{subsec:violating_pairs}.
Then, in \Cref{subsec:alg}, we describe our meta-algorithm.

\ifnum\fullversion=1 We \else In the full version of this paper~\cite{fullversion}, we \fi show that these conditions hold for all (non-trivial) threshold rules\ifnum\fullversion=1 ~(\Cref{subsec:thresh-rules})\fi, as well as a couple of additional rules\ifnum\fullversion=1 ~(\Cref{subsec:other-rules})\fi.

\subsection{The grid }\label{subsec:grid}

In this subsection we introduce the notion of a
one-dimensional ``grid'',
which will be a central building block of the meta algorithm (and its analysis). Recall that a configuration is a function $\sigma:\cycnums{n}\to\bitset$. A \emph{partial} configuration is a function $\sigma':\cycnums{n}\to\bitset \cup \bot$, which will serve us to denote restrictions of configurations to a subset of the locations.

Let $\Delta = \frac{\eps^2}{b_0}\cdot \min\{n,m\}$
where $b_0$ is a sufficiently large constant.
We assume for simplicity that $\Delta$ and $n/\Delta$ are both integers.
Let $ G \subseteq \cycnums{n} $ (the \textit{grid}) be the set of locations $\{j\cdot (n/\Delta)\}_{j=0}^{n/\Delta-1}$.

As we shall see in \Cref{subsec:alg}, our algorithm queries the tested environment on all grid locations and their $k$-neighborhoods at a specific time step $t_1$ (which will be set subsequently).

Let $\E_t[G]$ be the partial configuration that agrees with $\E_t$ on all locations in $\{\Gamma_k(g): g\in G\}$ and is $\bot$ elsewhere.

\begin{definition}\label{def:feasible}
	Given a time step $ t>0 $, we say that
	the partial configuration $\E_t[G]$ induced by the $k$-neighborhoods of the grid locations at time $t$
	is \textsf{feasible} with respect to 
	a rule $\rho$,
	if there exists  an environment $\E'$ that evolves according to $\rho$ and such that $\E'_t(i) = \E_t(i)$ for every $i \in \Gamma_k(G)$. We say in such a case that $\E'$ is a \textsf{feasible completion} of $\E_t[G]$ with respect to $\rho$. 
\end{definition}

\begin{definition}\label{def:grid_interval}
	Given a pair of grid locations $ g_1,g_2 \in G $, 
	a time step $ t $ and a subset $\mathcal{S} \subset \{0,1\}^{2k+1}$,
	if for every grid location $ g \in G \cap [g_1,g_2]$ it holds that $ \E_t(\Gamma_k(g))\in S$, then we say that the interval $ [g_1,g_2] $ is an \textsf{$S$ grid interval with respect to $\E_t$}.
	We say that $ [g_1,g_2] $ is a
	\textsf{maximal $\mathcal{S}$ grid interval with respect to $\E_t$},
	if both $ \E_t(g_1-\Delta)$ and $ \E_t(g_2+\Delta)$ do not belong to $\mathcal{S}$.
\end{definition}
In particular, we shall be interested in the case that $\mathcal{S}$ is $\F_\rho$ or $\bF_\rho$.
Note that a grid interval $ [g_1,g_2] $ contains all the locations between $ g_1 $ and $ g_2 $, and not just the grid locations.
Also note that if $\E_t(\Gamma_k(g)) \in S$ for every $g\in G$, then by \Cref{def:grid_interval}, there is no maximal $\mathcal{S}$ grid interval with respect to $\E_t$ (we shall deal with such cases separately).

\subsection{Violating Pairs}\label{subsec:violating_pairs}

Let $\rho$ be a fixed local rule that satisfies all the conditions stated in \Cref{subsec:conditions}.
Let $ t_1= \frac{b_1 \Delta}{\epsilon} $, where $b_1$ is a sufficiently large constant
and $\Delta$ is as defined in \Cref{subsec:grid}. Let $ t_2=t_1 + \Delta$.
We now define the concept of a violating pair $(t,i) \in \nums{m}\times \cycnums{n}$
with respect to $ \E_{t_1}$.
Generally speaking, these are pairs in the environment $ \E $ whose values are inconsistent with evolving according to the rule $ \rho $ given the values at the grid locations at time $ t_1 $.
The definition of a violating pair serves us later by allowing our algorithm to reject when it encounters one, which, as we prove, happens with high constant probability if $ \E $ is $ \epsilon $-far from evolving according to the rule $ \rho $.

\begin{figure}[htb!]
	\centerline{\mbox{\includegraphics[width=1\textwidth]{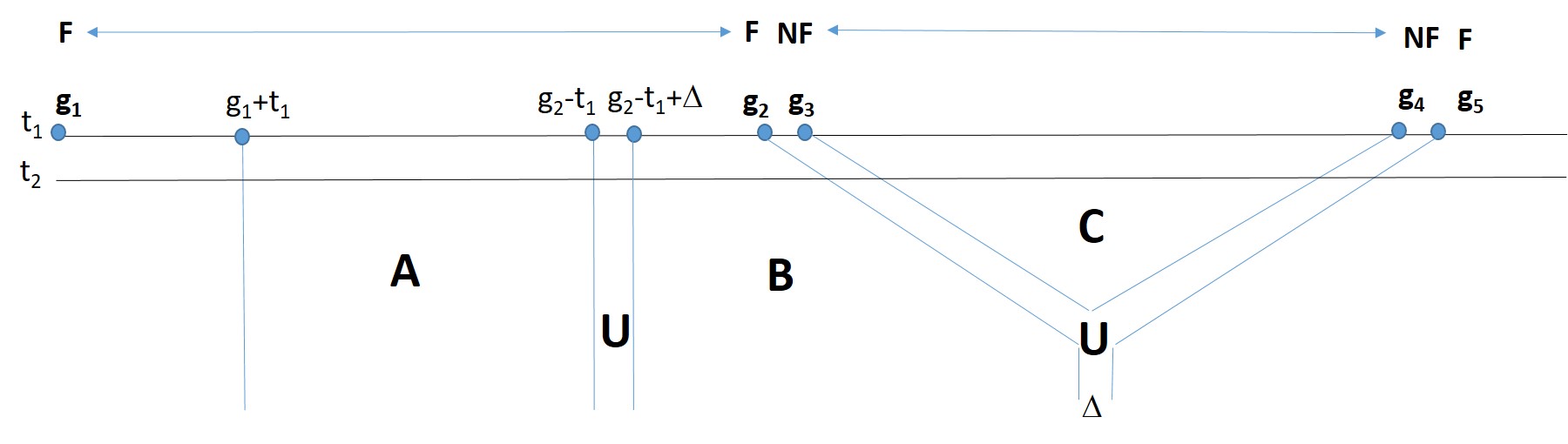}
	}}
	\caption{\small An illustration for the sets $A$, $B$, $C$, and $U$. Here $[g_1,g_2]$ is a maximal $\F_\rho$ grid interval, $g_3 = g_2+\Delta$, where $[g_3,g_4]$ is a maximal $\bF_\rho$ grid interval, and $g_5=g_4+\Delta$ is an endpoint of a maximal $\F_\rho$ grid-interval. The area marked by $A$ corresponds to pairs $(t,i)$ such that $t>t_2$
		and $i \in [g_1+ t_1,g_2-t_1]$. These pairs are supposed to be final.
		The area marked by $B$ corresponds to pairs $(t,i)$ such that $t>t_2$,
		$i \in [g_2-t_1+\Delta,g_2+(t-t_1)]$, and $\dist(g_2,i)< \dist(g_5,i) - \Delta$.
		These pairs are supposed to be final too.
		The area marked by $C$ corresponds to pairs  $(t,i)$ such that $t>t_2$,
		$i \in [g_3,g_4]$, and  $ (t,i) $ neither descends from
		$ (t_1,g_3+1) $ nor from  $ (t_1,g_4-1) $.
		These pairs are supposed to be non-final.
		Finally, the areas marked by $U$ correspond to pairs $(t,i)$ such that $t>t_2$ and one of the following holds: {\bf (1)} $i \in [g_2-t_1+1, g_2(i)-t_1+\Delta]$; {\bf (2)}
		$i \in [g_3,g_4]$ and either {\bf (a)} $ (t,i) $ descend from $ (t_1,g_3-\Delta) $ or $(t_1,g_4+\Delta)$
		and  $ |\dist(g_3, i) - \dist(g_4, i)| \le \Delta$, or {\bf (b)}
		$ (t,i) $ does not descend from either $ (t_1,g_3-\Delta)$ or $(t_1,g_4+\Delta)$ but it  descends from either $ (t_1,g_3) $ or  $ (t_1,g_4) $.}
	\label{fig:ABCU}
\end{figure}

We next define three disjoint sets of time-location pairs, denoted $ A $, $ B $ and $ C $, and for each of these three sets we state conditions under which the pair is considered to be a violating pair with respect to $ \E_{t_1}[G] $.
The proof that the three sets are pairwise disjoint appears in \Cref{subsec:observations},
and for an illustration, see \Cref{fig:ABCU}.

\smallskip
If $\E_{t_1}(\Gamma_k(g)) \in \F_\rho$ for every $g\in G$, then $A = \{(t,i): t_2 < t < m,\; i\in \cycnums{n}\}$. Otherwise,
$ A $ is the set of pairs $ (t,i) $ where $ t_2 < t < m $  and $ i \in \cycnums{n} $  such that there exists a maximal $\F_\rho$ grid interval $ [g_1(i), g_2(i)] $ with respect to $\E_{t_1}$ 
for which 
$i \in [g_1(i)+ t_1,g_2(i)-t_1]$.

\begin{definition}\label{def:A-violating}
	A pair $ (t,i) \in A $ is said to be a violating pair with respect to $ \E_{t_1}[G] $,
	if at least one of the following requirements  does not hold.
	~(1) $ \E_{t_2}(\Gamma_k(i)) \in \F_\rho $. ~(2) $ \E_t(\Gamma_k(i)) \in \F_\rho $.
	~(3) $ \E_t(i) = \frhor(\E_{t_2}(i), \parity(t-t_2), 0) $ where $ \frhor $ is the function referred to in \Cref{condition:final_prediction}.
\end{definition}

\smallskip
Let $ B $ be the set of pairs $(t,i)$,
where $ t_2 < t < m $ and $ i \in \cycnums{n} $   for which the following holds.
First, there exists a maximal $\F_\rho$ grid interval $ [g_1(i), g_2(i)] $ with respect to $\E_{t_1}$
such that either
$i \in [g_1(i) -(t-t_1),g_1(i)+t_1-\Delta-1]$ or $i \in [g_2(i)-t_1+\Delta+1,g_2(i)+(t-t_1)]$.
Second, for every other  maximal $\F_\rho$ grid interval $ [g'_1, g'_2] $ (with respect to $\E_{t_1}$),
if $i \in [g_1(i) -(t-t_1),g_1(i)+t_1-\Delta-1]$, then
$\dist(g_1(i), i) <\dist(g_2', i) - \Delta$, and if $i \in [g_2(i)-t_1+\Delta+1,g_2(i)+(t-t_1)]$,
then $\dist(g_2(i), i) <\dist(g_1', i) - \Delta$.

\begin{definition}\label{def:B-violating}
	A pair $ (t,i) \in B $ is said to be a violating pair with respect to $ \E_{t_1}[G] $, if at least one of the following requirements does not hold.
	~(1)  $\E_{t}(\Gamma_k(i)) \in \F_\rho $.
	~(2) Let $[g_1(i),g_2(i)]$ be the maximal $\F_\rho$ grid interval ensured by the definition of $B$ given $(t,i)$. Let $g(i)$ 
	be the grid location in $G \cap \left( [g_1(i),g_1(i)+t_1-\Delta] \cup [g_2(i),g_2(i)-t_1+\Delta] \right)$ that is closest to $i$ (if there are two such grid locations, then select the one closer to $g_1(i)$).
	Then $\E_t(i) = \frhor(\E_{t_1}(g(i)), \parity(t-t_1), \parity(\dist(i,g(i))))  $,
	where $ \frhor$ is the function referred to in \Cref{condition:final_prediction}.
\end{definition}

\smallskip
If $\E_{t_1}(\Gamma_k(g)) \in \bF_\rho$ for every $g\in G$, then $C = \{(t,i): t_2 < t < m,\; i\in \cycnums{n}\}$. Otherwise,
$ C $ is the set of pairs $ (t,i) $ where $ t_2 < t < m $ and $ i \in \cycnums{n} $ for which the following holds. First, there exists a maximal $\bF_\rho$ grid interval $ [g_1(i), g_2(i)] $ with respect to $\E_{t_1}$ such that
$i \in [g_1(i),g_2(i)]$.
Second, the pair $ (t,i) $ neither descends from the pair
$ (t_1,g_1(i)+1) $ nor from the pair $ (t_1,g_2(i)-1) $.

\begin{definition}\label{def:C-violating}
	A pair $ (t,i) \in C $ is said to be a violating pair with respect to $ \E_{t_1}$,
	if at least one of the following requirements does not hold.
	~(1) $ \E_{t}(\Gamma_k(i)) \in  \bF_\rho $.
	~(2) Let $g(i) \in G $  be a grid location satisfying $\dist(g(i),i)< \Delta$ (if there are two such grid locations, then select the one closer to $g_1(i)$).
	Then $ \E_t(\Gamma_k(i)) = \hrhor(\E_{t_1}(\Gamma_k(g(i))), \parity(t-t_1), \ddist(g(i),i))   $.
\end{definition}

Finally, we define the set $U$ of \emph{uncertain} pairs $(t,i)$, for which we cannot determine, given $\E_{t_1}[G]$ and the corresponding pairs $(t_2,i)$, whether they are violating or not.

\begin{definition}\label{def:U}
	The set $U$ consists of all pairs
	$ (t,i) \in \cycnums{n} \times \nums{m}$ such that $t>t_2$ and $(t,i) \notin  A \cup B \cup C $.
\end{definition}
In \Cref{subsec:observations} show that the number of pairs $ (t,i) $ belonging to the set $ U $ is relatively small, provided that $ \E_{t_1}[G] $ is feasible.

\subsection{The testing algorithm}\label{subsec:alg}

Recall that Let $\Delta = \frac{\eps^2}{b_0}\cdot \min\{n,m\}$,
$ t_1= \frac{b_1 \Delta}{\epsilon} $, and $t_2 = t_1 + \Delta$
(where $b_0$ and $b_1$ are constants that will be set in the analysis).

\begin{figure}[htb!]
	\fbox{\begin{minipage}[c]  {\textwidth}%
			\smallskip{}
			\textbf{Tester for evolution according to a rule $\rho$}
			\begin{enumerate}
				
				\item\label{step:reject_infeasible_grid} Query  $\E_{t_1}$ on all locations in $\Gamma_k(G)$. If $ \E_{t_1}[G] $ is infeasible with respect to $\rho$, reject.
				
				\item\label{step:select_random_pairs} Select uniformly at random $ \Theta(\frac{1}{\epsilon}) $ pairs $ (t,i) $ where $ i \in \cycnums{n} $ and $ t_2 < t < m $. \\ For each selected pair $ (t,i) $, query $ \E_{t}(\Gamma_k(i)) $ and $ \E_{t_2}(\Gamma_k(i)) $.
				
				\item\label{step:reject_violations} If some pair selected in Step~\ref{step:select_random_pairs} is a violating pair with respect to $\rho$, then reject. \\ Otherwise, accept.
				
			\end{enumerate}
			\vspace{3pt}
	\end{minipage}}
	\caption{The testing algorithm}\label{fig:alg}
\end{figure}

\setcounter{theorem}{2}

\begin{theorem}\label{thm:main}
	Let $\rho$ be any local rule that satisfies Conditions~\ref{condition:final}--\ref{condition:non-localB}.
	The algorithm described in \Cref{fig:alg} is a one-sided error non-adaptive testing algorithm for evolution according to $\rho$ whose query complexity is $O(\lceil n/m\rceil/\eps^2)$.
\end{theorem}
The bound on the query complexity of the algorithm follows from the fact that the number of queries performed in Step~\ref{step:reject_infeasible_grid} is $O(n/\Delta) = O(\lceil n/m \rceil/\eps^2)$ (recall that $k$ is a constant), and the number of queries performed in Step~\ref{step:reject_violations} is $O(1/\eps)$.
The correctness of the algorithm follows from the next two lemmas.
We prove \Cref{lemma:completeness} in \Cref{subsec:lem-E-obys} and \Cref{lemma:soundness} in \Cref{subsec:lem-E-far}.

\setcounter{lemma}{0}

\begin{lemma}[Completeness of the meta-algorithm]\label{lemma:completeness}
	Let $\rho$ be any local rule that satisfies Conditions~\ref{condition:final}--\ref{condition:non-localB}.
	If the environment $ \E $ evolves according to $\rho $,
	then the algorithm accepts with probability 1.
\end{lemma}

\begin{lemma}[Soundness of the meta-algorithm]\label{lemma:soundness}
	Let $\rho$ be any local rule that satisfies Conditions~\ref{condition:final}--\ref{condition:non-localB}.
	If the environment $ \E $ is $ \epsilon $-far from evolving according to $ \rho $, then the algorithm rejects with
	probability at least $2/3$.
\end{lemma}

\section{Observations and simple claims}\label{subsec:observations}

\setcounter{observation}{0}

In this subsection we present several observations and simple claims that will be used in our proofs of \Cref{lemma:completeness} and \Cref{lemma:soundness}.

The first two observations are directly implied by Conditions~\ref{condition:final} and~\ref{condition:infecting_neighbors}.

\begin{observation}\label{obs:F-F}
	Let $\rho$ be a local rule that satisfies Conditions~\ref{condition:final} and~\ref{condition:infecting_neighbors},
	$\E \in  \calE^\rho_{m,n}$ an environment that evolves according to $\rho$
	and  $(t,i) \in \nums{m}\times \cycnums{n}$. If $(t,i)$ has an ancestor $(t',i')$
	such that $\E_{t'}(\Gamma_k(i')) \in \F_\rho $, then $\E_{t}(\Gamma_k(i)) \in \F_\rho$.
\end{observation}
Note that \Cref{obs:F-F} implies that if $\E_{t}(\Gamma_k(i)) \in \bF_\rho$, then $\E_{t'}(\Gamma_k(i')) \in \bF_\rho$ for every ancestor $(t',i')$ of $(t,i)$.

\begin{observation}\label{obs:long_final_intervals}
	Let $\rho$ be a local rule that satisfies Conditions~\ref{condition:final} and~\ref{condition:infecting_neighbors},
	$\E \in  \calE^\rho_{m,n}$ an environment that evolves according to $\rho$
	and  $(t,i) \in \nums{m}\times \cycnums{n}$, $t \leq n/2$.
	If $\E_t(\Gamma_k(i))\in \F_\rho$, then the location $ i $ belongs to an interval whose size is at least $2t$  such that $\E'_t(\Gamma_k(j))\in \F_\rho$  for every location $j$ in this interval.
\end{observation}

\smallskip
The observation below directly follows from \Cref{obs:long_final_intervals}
(as well as the definition of the grid $G$ and Definitions~\ref{def:feasible} and~\ref{def:grid_interval}).
\begin{observation}\label{obs:max-F-grid-interval}
	Let $\rho$ be a local rule that satisfies Conditions~\ref{condition:final} and~\ref{condition:infecting_neighbors}.
	Suppose that $ \E_t[G] $ for $t \leq n/2$ is feasible with respect to $\rho$.
	Then 
	for every $[g_1,g_2]$ that is a maximal  $\F_\rho$ grid interval  with respect to $\E_{t}$,
	the number of locations in $[g_1,g_2]$ is at least $2t-\Delta$.
\end{observation}

The next observation follows directly from \Cref{obs:F-F}
(as well as the definition of $G$ and Definitions~\ref{def:feasible} and~\ref{def:grid_interval}).
\begin{observation}\label{obs:max-F-grid-interval-t2}
	Let $\rho$ be a local rule that satisfies Conditions~\ref{condition:final} and~\ref{condition:infecting_neighbors}.
	Suppose that $ \E_t[G] $ for $t \leq n/2$ is feasible with respect to $\rho$
	and let $g\in G$ be such that $\E_t(\Gamma_k(g)) \in \F_\rho$.
	If $ \E' $ is a feasible completion of $ \E_t[G]$ (with respect to $\rho$),
	then $\E'_{t'}(\Gamma_k(i)) \in \F_\rho$ for every
	$t' \geq t+\Delta$ and
	$i \in [g-\Delta,g+\Delta]$.
\end{observation}

\Cref{claim:max-bF-grid-interval}, stated next, also deals with feasible completions.
\begin{claim}\label{claim:max-bF-grid-interval}
	Let $\rho$ be a local rule that satisfies Conditions~\ref{condition:final} and~\ref{condition:infecting_neighbors}.
	Suppose that $ \E_t[G] $ is feasible with respect to $\rho$ for $t\geq\Delta$ and let $g\in G$ be such that both $\E_t(\Gamma_k(g)) \in \bF_\rho $ and $\E_t(\Gamma_k(g+\Delta)) \in \bF_\rho $.
	If $ \E' $ is a feasible completion of $ \E_t[G]$ (with respect to $\rho$),
	then $\E'_{t}(\Gamma_k(i)) \in \bF_\rho $ for every  
	$i\in [g,g+\Delta]$.
\end{claim}

\ifnum\fullversion=1
\begin{proof}
	Assume, contrary to the claim, that there exists some 
	$i\in [g,g+\Delta]$
	such that
	$\E'_t(\Gamma_k(i)) \in F_\rho$ (where $\E'$ is a feasible completion of $ \E_t[G]$ (with respect to $\rho$)).
	By Conditions~\ref{condition:final} and~\ref{condition:infecting_neighbors},
	there must be some location $j$ such that $(t,i)$ descends from $(0,j)$ and
	$\E'_0(\Gamma_k(j)) \in \F_\rho$. But this would imply
	(once again by \Cref{condition:infecting_neighbors}),
	that $\E'_t(\Gamma_k(g')) \in F_\rho$ for $g'=g$ or $g' =g+\Delta$, and we have reached a contradiction.
\end{proof}
\fi

\smallskip
Recall the definitions of the sets $A$, $B$ and $C$ from \Cref{subsec:violating_pairs}.
\begin{claim}
	The sets $ A $, $ B $, and $ C $ are pairwise disjoint.
\end{claim}

\ifnum\fullversion=1
\begin{proof}
	We first show that $ A \cap B = \emptyset $.
	Suppose by way of contradiction that there exists a pair $ (t,i) \in A \cap B $.
	In order for a pair $ (t,i) $ to belong to the set $ A $, it must hold that $ i \in [g_1^A+t_1,g_2^A-t_1] $ where $ [g_1^A,g_2^A] $ is a maximal $\F_\rho$ grid interval with respect to $\E_{t_1}$.
	In order for a pair $ (t,i) \in A $ to also belong to the set $ B $, it must hold that $i \in [g_1^B -(t-t_1),g_1^B+t_1-\Delta-1]$ or $i \in [g_2^B-t_1+\Delta+1,g_2^B+(t-t_1)]$ where $ [g_1^B,g_2^B] $ is a maximal $\F_\rho$ grid interval with respect to $\E_{t_1}$.
	
	Since $ (t,i) \in A $, then, in particular, the location $ i $ must belong to a maximal $\F_\rho$ grid interval with respect to $\E_{t_1}$.
	In the cases where $ i \in [g_1^B -(t-t_1),g^B_1-1] $ and $ [g_2^B+1,g_2^B+(t-t_1)] $, the location $ i $ cannot belong to a maximal $ \F_\rho $ grid interval.
	The reason is that, first, by the definition of the set $ B $, for every other maximal $\F_\rho$ grid interval $ [g'_1, g'_2] \ne [g_1^B,g_2^B] $ (with respect to $\E_{t_1}$), if $i \in [g_1(i) -(t-t_1),g_1(i)+t_1-\Delta-1]$, then $\dist(g_1(i), i) < \dist(g_2', i) - \Delta$, and if $i \in [g_2(i)-t_1+\Delta+1,g_2(i)+(t-t_1)]$, then $\dist(g_2(i), i) <\dist(g_1', i) - \Delta$.
	Hence, in the cases in which $ i \in [g_1^B -(t-t_1),g^B_1-1] $ or $ [g_2^B+1,g_2^B+(t-t_1)] $, if the location $ i $ were inside a maximal $ \F_\rho $ grid interval, then that interval would be adjacent to the interval $ [g_1^B,g_2^B] $.
	However, this is impossible, since between each pair of maximal $ \F_\rho $ grid intervals there is a maximal $\bF_\rho$ grid interval with respect to $ \E_{t_1} $.
	Therefore, it must hold that either $i \in [g_1^B,g_1^B+t_1-\Delta-1]$ or $i \in [g_2^B-t_1+\Delta+1,g_2^B]$.
	If $ [g_1^A,g_2^A] \ne [g_1^B,g_2^B] $, then it clearly cannot hold that the location $ i $ also belongs to $ [g_1^A+t_1,g_2^A-t_1] $ (because the maximal $\F_\rho$ grid intervals are disjoint with respect to each other).
	Also, if $ [g_1^A,g_2^A] = [g_1^B,g_2^B] $, it also cannot hold that $ i \in [g_1^A+t_1,g_2^A-t_1] $ because in this case all the locations belonging the interval $ [g_1^B,g_1^B+t_1-\Delta-1] $ are to the left of all the locations belonging to the interval $ [g_1^A+t_1,g_2^A-t_1] $.
	Similarly, all the locations belonging the interval $ [g_2^B-t_1+\Delta+1,g_2^B] $ are to the right of all the locations belonging to the interval $ [g_1^A+t_1,g_2^A-t_1] $.
	Hence, if $ (t,i) \in A $ then $ (t,i) \notin B $, and therefore $ A \cap B = \emptyset $.
	
	We now show that $ B \cap C = \emptyset $.
	Suppose by way of contradiction that there exists a pair $ (t,i) \in B \cap C $.
	Since $ (t,i) \in C $, it must hold that $ i \in [g_1^C,g_2^C] $ where $ [g_1^C,g_2^C] $ is a maximal $ \bF_\rho $ grid interval with respect to $ \E_{t_1} $ and that $ (t,i) $ neither descends from the pair $ (t_1,g_1^C+1) $ nor from the pair $ (t_1,g_2^C-1) $.
	For $ (t,i) $ to also belong to the set $ B $, it must hold that $i \in [g_1^B -(t-t_1),g_1^B+t_1-\Delta-1]$ or $i \in [g_2^B-t_1+\Delta+1,g_2^B+(t-t_1)]$ where $ [g_1^B,g_2^B] $ is a maximal $\F_\rho$ grid interval with respect to $\E_{t_1}$.
	However, since $ (t,i) \in C $, and therefore, in particular, the location $ i $ must belong to a maximal $\bF_\rho$ grid interval with respect to $\E_{t_1}$, the cases where $i \in [g_1^B,g_1^B+t_1-\Delta-1]$ and $i \in [g_2^B-t_1+\Delta+1,g_2^B]$ cannot hold.
	Therefore, it must hold that either $ i \in [g_1^B -(t-t_1),g_1-1] $ or $ i \in [g_2^B+1,g_2^B+(t-t_1)] $.
	If $ [g_1^C,g_2^C] $ and $ [g_1^B,g_2^B] $ are not adjacent maximal grid intervals (adjacent in the sense that they are at a distance of $ \Delta $ from each other), then the location $ i $ cannot also belong to $ [g_1^C,g_2^C] $.
	Hence, we can assume without loss of generality that $ i \in [g_2^B+1,g_2^B+(t-t_1)] $ and $ g_1^C=g_2^B+\Delta $.
	Thus, $ \dist(i,g_1^C+1) = \dist(i,g_2^B+\Delta+1) \le t - t_1 $.
	That is, the pair $ (t,i) $ descends from the pair $ (t_1,g_1^C+1) $, in contradiction to $ (t,i) \in C $.
	Therefore, no pair $ (t,i) $ cannot belong to both $ B $ and $ C $.
	
	Finally, to see that $ A \cap C = \emptyset $, note that in order for a pair $ (t,i) $ to belong to the set $ A $ it must belong to a maximal $\F_\rho$ grid interval with respect to $\E_{t_1}$, and in order for a pair $ (t,i) $ to belong to the set $ C $ it must belong to a maximal $ \bF_\rho $ grid interval with respect to $ \E_{t_1} $.
	Therefore, no $ (t,i) $ pair can belong to both $ A $ and $ C $, and hence $ A \cap C = \emptyset $.
\end{proof}
\fi

\smallskip
In the last claim of this subsection, we bound the size of the set $U$ of uncertain pairs (as defined in \Cref{def:U}).

\begin{claim}\label{claim:U_is_small}
	If $ \E_{t_1}[G] $ is feasible (with respect to $\rho$), then $ |U| \leq \frac{5\eps}{b_1} m n $
	(where $b_1$ is the constant in the setting of $t_1 = \frac{b_1\Delta}{\eps}$).
\end{claim}
We note that \Cref{claim:U_is_small} does not depend on the setting of $\Delta$, but only on the definition of $t_1$ as a function of $\Delta$ (as well as the definition of the grid $G$, which, too is defined based on $\Delta$, and in turn is used to determine $U$).

\ifnum\fullversion=1
\begin{proof}
	First note that if $\E_{t_1}(\Gamma_k(g)) \in \F_\rho$ for every $g \in G$ or
	$\E_{t_1}(\Gamma_k(g)) \in \bF_\rho$ for every $g \in G$, then $U$ is empty.
	Hence, we assume from this point on that neither is the case, so that there is at least one (non-empty) maximal $\F_\rho$ grid interval and at least one (non-empty) maximal $\bF_\rho$ grid interval.
	
	By the definition of $U$ and the sets $A$, $B$, and $C$, a pair $ (t,i) $ belongs to the set $ U $ if $t> t_2$ and
	one of the following holds.
	\begin{enumerate}
		\item There exists a maximal $\F_\rho$ grid interval $ [g_1(i), g_2(i)] $, such that either
		$i\in [g_1(i)+t_1-\Delta, g_1(i)+t_1-1]$
		or
		$i \in [g_2(i)-t_1+1, g_2(i)-t_1+\Delta]$.
		\item There exists a maximal $\bF_\rho$ grid interval $ [g_1(i), g_2(i)] $
		with respect to $\E_{t_1}$
		where 
		$i \in [g_1(i),g_2(i)]$. Furthermore, if the pair $ (t,i) $ descends from at least one of the pairs $ (t_1,g_1(i)-\Delta) $ or $ (t_1,g_2(i)+\Delta) $, then $ |\dist(g_1(i), i) - \dist(g_2(i), i)| \le \Delta$. Otherwise (the pair $ (t,i) $ does not descend from either $ (t_1,g_1(i)-\Delta) $ or $(t_1,g_2(i)+\Delta)$), the pair $ (t,i) $ descends from either the pair $ (t_1,g_1(i)) $ or from the pair $ (t_1,g_2(i)) $.
	\end{enumerate}
	By \Cref{obs:max-F-grid-interval}, the length of each maximal $\F_\rho$ grid interval at time $t_1$ is at least $t_1$. Therefore, the  number of $\F_\rho$ grid intervals is at most $\frac{n}{t_1} = \frac{\eps n}{b_1\Delta}$. This implies that the total number of pairs $(t,i)\in U$ of the first type described above, is at most $2\Delta \cdot \frac{\eps n}{b_1\Delta} \cdot m = \frac{2\eps}{b_1} m n $.
	Since between every two maximal $\bF_\rho$ grid intervals  there is a maximal $\F$ grid interval,
	the number of maximal $\bF_\rho$ grid intervals is also upper bounded by $\frac{\eps n}{b_1\Delta}$.
	We shall show that for each maximal $\bF$ grid interval $[g_1,g_2]$, the number of pairs in $U$ that descend from either $ (t_1,g_1) $ or $ (t_1,g_2) $ is at most $3\Delta m$, and the current claim follows.
	
	Consider any fixed maximal $\bF_\rho$ grid interval $[g_1,g_2]$. The number of pairs
	$ (t,i) $, 
	where $i \in [g_1,g_2]$,
	that descend from at least one of the pairs $ (t_1,g_1) $ or $ (t_1,g_2) $, and for which $ |\dist(g_1, i) - \dist(g_2, i)| \le \Delta$  is at most $\Delta m$. This upper bound actually follows by using only the condition  $ |\dist(g_1, i) - \dist(g_2, i)| \le \Delta$.
	The number of pairs  $ (t,i) $, where $i \in [g_1,g_2]$, 
	that do not descend from either $ (t_1,g_1-\Delta) $ or $(t_1,g_2+\Delta)$, but do descend from either  $ (t_1,g_1) $ or $ (t_1,g_2) $ is at most $2\Delta m$.
	This follows directly from the definition of descending pairs.
\end{proof}
\fi

\section{Proof of {\Cref{lemma:completeness}}: \nameref{lemma:completeness}}\label{subsec:lem-E-obys}
Let $\rho$ be any local rule that satisfies Conditions~\ref{condition:final}--\ref{condition:non-localB}
(where in this proof we  do not make use of Conditions~\ref{condition:non-localA} and~\ref{condition:non-localB}, which are provided in the next subsection),
and	let $ \E \in \calE^\rho_{m,n} $ be a dynamic environment that evolves according to $ \rho $.
The only steps in which our algorithm may reject are Step~\ref{step:reject_infeasible_grid} and Step~\ref{step:reject_violations}.  The grid is feasible by definition, and hence the algorithm does not reject in Step~\ref{step:reject_infeasible_grid}. To show that it also does not reject in Step~\ref{step:reject_violations},
we show that there are no violating pairs with respect to  $ \E_{t_1}[G] $.
Recall that each violating pair belongs to one of the three sets $A$, $B$, or $C$ (as defined in \Cref{subsec:violating_pairs}).
Specifically, we next show that in each of the three cases ($ (t,i) \in A $, $ (t,i) \in B $, and $ (t,i) \in C $), the requirements (specified in \Cref{subsec:violating_pairs}) for $ (t,i) $ being a non-violating pair hold.
In what follows, if we say that a pair $(t,i)$ is final (similarly, non-final), then we mean with respect to $\E$, and when we refer to maximal grid intervals, it is always with respect to $\E_{t_1}$,
and violations are always with respect to $ \E_{t_1}[G] $.


\subparagraph{Pairs {\boldmath{$ (t,i) \in A $}}.}
By the definition of $A$,
$t > t_2$  and
there exists a grid location $g(i)\in G$ such that $\dist(i,g(i)) \leq \Delta$ and $\E_{t_1}(\Gamma_k(g(i)))\in \F_\rho$
(this holds both in the case that $\E_{t_1}(\Gamma_k(g))\in \F_\rho$ for every $g\in G$ and in the case that
there exists a maximal $\F_\rho$ grid interval $[g_1(i),g_2(i)]$ such that
$i \in [g_1(i)+ t_1,g_2(i)-t_1]$.)
By \Cref{obs:max-F-grid-interval-t2}, both $\E_{t_2}(\Gamma_k(i)) \in \F_\rho$ and
$\E_t(\Gamma_k(i)) \in \F_\rho$.
Turning to the third requirement, by \Cref{condition:final_prediction}, applied with $t'=t_2$ and $i'=i$, we get that
$\E_t(i) = \frhor(\E_{t_2}(i), \parity(t-t_2), 0) $. Therefore, all three requirements on pairs in $A$ hold, and hence $(t,i)$ is not a violating pair.

\subparagraph{Pairs {\boldmath{$ (t,i) \in B $}.}}
By the definition of $B$, $t > t_2$ and  there exists a maximal $\F_\rho$ grid interval $ [g_1(i), g_2(i)] $ with respect to $\E_{t_1}$ such that either
$i \in [g_1(i) -(t-t_1),g_1(i)+t_1-\Delta-1]$ or $i \in [g_2(i)-t_1+\Delta+1,g_2(i)+(t-t_1)]$.
Furthermore,  for every other  maximal $\F_\rho$ grid interval $ [g'_1, g'_2] $ (with respect to $\E_{t_1}$),
if $i \in [g_1(i) -(t-t_1),g_1(i)+t_1-\Delta-1]$, then
$\dist(g_1(i), i) <\dist(g_2', i) - \Delta$, and if $i \in [g_2(i)-t_1+\Delta+1,g_2(i)+(t-t_1)]$,
then $\dist(g_2(i), i) <\dist(g_1', i) - \Delta$.
Let $g(i)$ 
be the  grid location closest to $i$ in
$G \cap \left( [g_1(i),g_1(i)+t_1-\Delta] \cup [g_2(i)-t_1+\Delta,g_2(i)] \right)$
(as defined in the second requirement concerning (non-)violating pairs $(t,i)\in B$).

We claim that $(t,i)$ descends from $(0,g(i))$.
To see why, first consider the case in which $ i \in [g_1(i)-(t-t_1),g_1(i)] \cup [g_2(i),g_2(i)+(t-t_1)] $.
In this case, either $ g(i)=g_1(i) $ or $ g(i)=g_2(i) $, which means that $ \dist(i,g(i)) \le t-t_1 \le t $.
Second, consider the case in which $ i \in [g_1(i),g_1(i)+t_1-\Delta-1] \cup [g_2(i)-t_1+\Delta+1,g_2(i)] $.
In this case, the grid location closest to $ i $ in $ G \cap \left( [g_1(i),g_1(i)+t_1-\Delta] \cup [g_2(i)-t_1+\Delta,g_2(i)] \right) $ is within a distance of at most $ \Delta $ from the location $ i $.
Hence, $ \dist(i,g(i)) \le \Delta \le t $.
Therefore, in any case, $ \dist(i,g(i)) \le t $, and thus the pair $ (t,i) $ descends from the pair $ (0,g(i)) $.

Assume (without loss of generality) that $g(i) \in [g_2(i)-t_1+\Delta,g_2(i)]$.
Since $\E_{t_1}(\Gamma_k(g_2(i) + \Delta))\in \bF_\rho$ (as $[g_1(i),g_2(i)]$ is a maximal final grid interval),
we know (by \Cref{obs:F-F}) that
$\E_0(\Gamma_k(j)) \in \bF_\rho$ for every $j \in [g_2(i)+\Delta - t_1,g_2(i)+\Delta + t_1 ]$.
However, since $\E_{t_1}(\Gamma_k(g_2(i))\in \F_\rho$, there must be some location $\ell \in [g_2(i)-t_1, g_2(i)+\Delta - t_1-1]$
such that $\E_0(\Gamma_k(\ell)) \in \F_\rho$. Among the locations $\ell$ that satisfy these conditions, let $\ell^*$ be the one that minimizes $\dist(\ell,g_2(i)+\Delta - t_1)$, so that for every $\ell' \in [\ell^*+1, g_2(i)+\Delta - t_1]$ we have that $\E_0(\Gamma_k(\ell')) \in \bF_\rho$.
Hence, for every $ i'' \ne g(i) $ satisfying $ \dist(g(i),i'') \le \dist(g(i),\ell^*) $ it holds that $ \E_0(i'') \in \bF_\rho $.
Additionally, since $ g(i) \in [g_2(i)-t_1+\Delta,g_2(i)] $ and $ \ell^* \in [g_2(i)-t_1, g_2(i)+\Delta - t_1-1] $, it must hold that $ \dist(g(i),\ell^*) \le t_1 $, which means that the pair $ (t_1,g(i)) $ descends from the pair $ (0,\ell^*) $.
Also, both $ \E_{t_1}(g(i)) \in \F_\rho $ and $ \E_0(\ell^*) \in \F_\rho $.
Thus, we can apply \Cref{condition:final_prediction} for the two pairs $ (0,\ell^*) $ and $ (t_1,g(i)) $ to get that
$ \E_{t_1}(g(i)) = \frhor(\E_{0}(\ell^*), \parity(t_1), \parity(\dist(\ell^*,g(i)))) $.

Since the pair $ (t,i) $ descends from the pair $ (t_1, g(i)) $, and the pair $ (t_1, g(i)) $ descends from the pair $ (0, \ell^*) $, it holds that the pair $ (t,i) $ must also descend from the pair $ (0,\ell^*) $.
Additionally, both both $ \E_{t}(i) \in \F_\rho $ and $ \E_0(\ell^*) \in \F_\rho $.
Also, by the second requirement on $ (t,i) $, involving other maximal $ \F_\rho $ grid intervals $ [g'_1, g'_2] $, for every $ i'' \ne i $ satisfying $ \dist(i,i'') \le \dist(i,\ell^*) $ it holds that $ \E_0(i'') \in \bF_\rho $.
Thus, we can apply \Cref{condition:final_prediction} for the two pairs $ (0,\ell^*) $ and $ (t,i) $ to get that
$ \E_{t}(i) = \frhor(\E_{0}(\ell^*), \parity(t), \parity(\dist(\ell^*,i))) $.
But then, since $\frhor$  is the XOR of its first argument and a subset of the other two,
and $\parity(t-t_1) = \parity(t_1) \oplus \parity(t)$ as well as
$ \parity(\dist(g(i),i))) =  \parity(\dist(\ell^*,g(i))) \oplus  \parity(\dist(\ell^*,i))$,
we get that
$ \E_{t}(i) = \frhor(\E_{t_1}(g(i)), \parity(t-t_1), \parity(\dist(g(i),i))) $.

\subparagraph{Pairs {\boldmath{$ (t,i) \in C $}}.}
There are two cases (where in both $t> t_2$).
The first is that $\E_{t_1}(\Gamma_k(g)) \in \bF_\rho$ for every $g\in G$ (so that $i$ may be any location in $\cycnums{n}$).
In the second case there exists a maximal 
$\bF_\rho$ grid interval $ [g_1(i), g_2(i)] $ 
such that  
$i \in [g_1(i),g_2(i)]$,
and  $(t,i) $ does not descend from either
$(t_1,g_1(i)-1) $ or  $(t_1,g_2(i)+1) $, which implies that for every $ j \in \cycnums{n} $, if the pair $ (t,i) $ descends from $ (t_1,j) $, then $ j \in [g_1(i),g_2(i)] $.
In both cases, by \Cref{claim:max-bF-grid-interval},
all ancestors $(t_1,j)$ of $(t,i)$ satisfy $\E_{t_1}(\Gamma_k(j)) \in \bF_\rho $.
By \Cref{obs:F-F} this implies that
$ \E_{t}(\Gamma_k(i)) \in  \bF_\rho $, so that the first requirement is met.
As for the second requirement, since
the grid location $g(i)$ defined in the second requirement is such that $(t_1,g(i))$ is an ancestor of $(t,i)$
(and $\E_{t_1}(\Gamma_k(g(i))) \in \bF_\rho $),
we can apply \Cref{condition:noninf_prediction} (with $t'=t_1$ and $i'=g$) to get that
$  \E_t(\Gamma_k(i))  = \hrhor(\E_{t_1}(\Gamma_k(g(i))), \parity(t-t_1), \ddist(g(i),i) $, as required.

We've shown that under the premise of the lemma, there is no pair $(t,i) \in A\cup B \cup C$ that is a violating pair.
Thus, our algorithm cannot reject at Step~\ref{step:reject_violations}.

\section{Proof of {\Cref{lemma:soundness}}: \nameref{lemma:soundness}}\label{subsec:lem-E-far}

Let $\E$ be any environment that is $ \epsilon $-far from evolving according to $\rho$,
where  $\rho$ is a local rule that satisfies Conditions~\ref{condition:final}--\ref{condition:non-localB}.
If $ \E_{t_1}[G] $ is infeasible with respect to $\rho$, then the algorithm rejects (in Step~\ref{step:reject_infeasible_grid}).
Hence, we assume from now on that $ \E_{t_1}[G] $ is feasible.

We claim that the number of violating pairs with respect to $ \E_{t_1}[G] $ is at least
$\frac{\eps}{b_2}mn$, where $b_2 > 1$ is a constant. \Cref{lemma:soundness} follows, since the algorithm selects
$s=\Theta(1/\eps)$ pairs (in Step~\ref{step:select_random_pairs}), and rejects if any of them is found to be a violating pair (in Step~\ref{step:reject_violations}). Hence, the probability that the algorithm rejects is at least $1-(1-\eps/b_2)^s$, which is at least $2/3$ for $s \geq 2b_2/\eps$.

Suppose by way of contradiction that there are less than $\frac{\eps}{b_2}mn$ violating pairs.
We show how, based on $\E$ (to be precise, $\E_{t_1}[G]$ and $\E_{t_2}$) we can define an environment $\E'$ for which the following holds. First, $\E'$ evolves according to $\rho$. Second, $\E'$ differs from $\E$ on at most $\eps m n$ pairs $(t,i) \in \cycnums{n}\times \nums{m}$. But this contradicts the premise that $ \E $ is $ \epsilon $-far from evolving according to $\rho$. Details follow in the next subsections.

We first provide all details (in \Cref{subsubsec:E-prime} and \Cref{subsubsec:E-E-prime}) under the assumption that there exist grid locations $g\in G$ for which $\E_{t_1}(\Gamma_k(g))\in \F_\rho$ as well as grid locations $g'\in G$ for which $\E_{t_1}(\Gamma_k(g'))\in \bF_\rho$.
We discuss (in \ifnum\fullversion=1 \Cref{subsubsec:homogeneous}\else the full version of this paper~\cite{fullversion}\fi ) the two special cases for which either $\E_{t_1}(\Gamma_k(g))\in \F_\rho$ for every $g\in G$ or  $\E_{t_1}(\Gamma_k(g))\in \bF_\rho$ for every $g\in G$, which we refer to as the \emph{homogeneous} cases.

\subsection{The definition of $\E'$}\label{subsubsec:E-prime}
To construct the dynamic environment $ \E' $, we define its initial configuration $ \E'_0 $, and then apply the local rule $ \rho $ for $ m -1$ steps. Hence, $\E'$ evolves according to $\rho$ by construction.
The initial configuration  $ \E'_0 $ is defined with respect to a configuration  $ \sigma $ on which we perform several
transformations to obtain $ \E'_0 $.
We define the configuration $ \sigma $ by specifying the value of $ \sigma(i) $ for each location $ i \in \cycnums{n} $
as explained next. In what follows, whenever we refer to maximal $\bF_\rho$ grid intervals (similarly, maximal $\F_\rho$ grid intervals), it is with respect to $\E_{t_1}$.

We shall make use of a function
$\hrhol: \bF_\rho \times \bitset \times \cycnums{n}$
(based on $\hrhor$ -- see \Cref{condition:noninf_prediction}).
Recall that by \Cref{condition:noninf_prediction}, 
for each fixed $\tau \in \bF_\rho$, $\beta \in \bitset$ and  $\ell \in \cycnums{n}$, there exists a unique $\tau'$ such that  $\hrhor(\tau',\beta,\ell) = \tau$.

\begin{definition}\label{def:h-lar}
	For any $\tau \in \bF_\rho$, $\beta \in \bitset$ and $\ell \in \cycnums{n}$,
	$\hrhol(\tau,\beta,\ell)$ equals the (unique) pattern $\tau'$ for which
	$\hrhor(\tau',\beta,\ell) = \tau$.
\end{definition}

We also make the following observation, based on \Cref{condition:final_prediction}, by which $\frhor$ is the XOR of its first argument and a subset of the other two.
\begin{observation}\label{obs:f-lar}
	\sloppy
	For any $\beta_1,\beta_2,\beta_3 \in \bitset$,
	if $\frhor(\beta_1,\beta_2,\beta_3) = \beta_1'$, then
	$\frhor(\beta_1',\beta_2,\beta_3) = \beta_1$.
	Furthermore, for any $\beta_2',\beta_3' \in \bitset$,
	$\frhor(\frhor(\beta_1,\beta_2,\beta_3),\beta'_2,\beta'_3) = \frhor(\beta_1,\beta_2\oplus \beta_2',\beta_3\oplus\beta_3')$, and in particular,
	$\frhor(\frhor(\beta_1,\beta_2,\beta_3),\beta_2,\beta_3) = \beta_1$.
\end{observation}

For each maximal $ \bF_\rho $ grid interval  $ [g_1,g_2] $,
let 
$J(g_1,g_2) = [g_1-t_1-k,g_2+t_1+k]$
and let $J$ be the union over all such sets.
We also define 
$J_1(g_1,g_2) = [g_1-t_1,g_2+t_1]$
(for each $ \bF_\rho $ grid interval  $ [g_1,g_2] $), and let $J_1\subset J$ be the union over all such sets.

We first establish two simple claims.

\begin{claim}\label{claim:disj-J}
	Let $\rho$ be any local rule that satisfies Conditions~\ref{condition:final}--\ref{condition:noninf_prediction}.
	For every two maximal $ \bF_\rho $ grid interval  $ [g_1,g_2] $ and $[g_1',g_2']$, we have that
	$J(g_1,g_2)\cap J(g'_1,g'_2) = \emptyset$.
\end{claim}

\ifnum\fullversion=1
\begin{proof}
	Since $ [g_1,g_2] $ and $[g_1',g_2']$ are maximal $ \bF_\rho $ grid intervals, we know that
	$g_1 - \Delta$, $g_2 + \Delta$, $g_1' - \Delta$ and $g_2'+\Delta$ are endpoints of maximal $\F_\rho$ grid intervals
	(with respect to $\E_{t_1}[G]$). By \Cref{obs:max-F-grid-interval-t2} these maximal $\F_\rho$ grid intervals are of size at least $2t_1$ each. As $k$ is a constant while $\Delta = \Theta(\eps^2 \min\{n,m\})$, we get that
	$ [g_1 - t_1 - k, g_2 + t_1 + k] $ must be disjoint from $ [g'_1 - t_1 - k, g'_2 + t_1 + k] $.
\end{proof}
\fi

\begin{claim}\label{claim:max-bF-zero}
	Let $\rho$ be any local rule that satisfies Conditions~\ref{condition:final}--\ref{condition:noninf_prediction}.
	Let $\E''$ be any environment that is a feasible extension of $\E_{t_1}[G]$ with respect to $\rho$, and let $[g_1,g_2]$ be a maximal $ \bF_\rho $ grid interval (with respect to $\E_{t_1}[G]$).
	Then $\E''_0(\Gamma_k(i)) \in \bF_\rho$ for every 
	$i \in J_1(g_1,g_2)$.
	Furthermore,  $\E''_0(\Gamma_k(i)) = \hrhol(\E_{t_1}(\Gamma_k(g)), \parity(t_1), \ddist(i,g))$
	for any $g \in G\cap [g_1,g_2]$ and every ancestor $(0,i)$ of $(t_1,g)$. 
\end{claim}

\ifnum\fullversion=1
\begin{proof}
	The first part of the claim follows 
	from \Cref{condition:final} and \Cref{condition:infecting_neighbors}. Namely, assume, contrary to the claim, that
	$\E''_0(\Gamma_k(i)) \in \F_\rho$ for some $g_1-t_1 \leq i \leq g_2+t_2$. But then,
	by  \Cref{condition:final} and \Cref{condition:infecting_neighbors},  there would be at least one grid location $g \in G\cap [g_1,g_2]$ such that $\E''_{t_1}(\Gamma_k(g)) \in \F_\rho$. This stands  in contradiction to $[g_1,g_2]$ being a maximal $ \bF_\rho $ grid interval
	with respect to $\E_{t_1}[G]$, and $\E''$ being a feasible extension of $\E_{t_1}[G]$.
	
	\sloppy
	As for the second part of the claim, by \Cref{condition:noninf_prediction} (which can be applied given the first part of the claim), for any $i$ and $g$ as defined in the claim,
	$\E''_{t_1}(\Gamma_k(g)) = \hrhor(\E''_0(\Gamma_k(i)) , \parity(t_1), \ddist(i,g))$. But then, by the definition of $\hrhol$ and that premise that $\E''$ is a feasible extension of $\E_{t_1}[G]$,
	$\E''_0(\Gamma_k(i)) = \hrhol(\E_{t_1}(\Gamma_k(g)), \parity(t_1), \ddist(i,g))$, as claimed.
\end{proof}
\fi

Observe that \Cref{claim:max-bF-zero} implies that
$\E''_0$ is \emph{uniquely} determined by $\E_{t_1}[G]$ on all location in $J$ for every $\E''$ that is a feasible  extension of $\E_{t_1}[G]$ (with respect to $\rho$).
Based on this observation,
we start by setting the locations of $\sigma$ that belong to $J$ as in  such $\E''_0$.
In particular we have
that $\sigma(\Gamma_k(i)) \in \bF_\rho$ for every $i \in J_1$, and furthermore,

\ifnum\fullversion=0
\vspace{-3.5ex}
\fi
\begin{equation}\label{eq:sigma-J}
	\begin{split}
		\forall i\in J_1, g\in G & \text{ s.t. } (t_1,g) \text{ descends from } (0,i) \text{ and,} \\
		&\sigma(\Gamma_k(i)) = \hrhol(\E_{t_1}(\Gamma_k(g)), \parity(t_1), \ddist(i,g)) \;.
	\end{split}
\end{equation}

Turning to the locations not yet set in $\sigma$, for each location $ i \in \cycnums{n} \setminus J $,
\ifnum\fullversion=1
\[ \sigma(i) = \frhor(\E_{t_2}(i), \parity(t_2), 0) \;.\]
\else
$\sigma(i) = \frhor(\E_{t_2}(i), \parity(t_2), 0)$.
\fi
Note that by \Cref{obs:f-lar},
$\frhor(\sigma(i), \parity(t_2), 0) = \E_{t_2}(i)$.

\smallskip
We next explain how we modify $ \sigma $ so as to obtain $\E'_0$
using \Cref{condition:non-localA} and \Cref{condition:non-localB}.
The modifications are performed (strictly) within the following set of intervals $\calS$.

\ifnum\fullversion=0
\vspace{-3.5ex}
\fi
\begin{equation}\label{eq:calS}
	\mathcal{S} = \set{\big[a=g_1 - \Delta + t_1, b=g_2 + \Delta - t_1\big] \medspace{ }  : 
		\medspace{ }
		\text{$ [g_1,g_2] $ is a maximal $ \F_\rho $ grid interval }} \;.
\end{equation}

The intervals in $\mathcal{S}$ are clearly disjoint (as each is a sub-interval of a different maximal $ \F_\rho $ grid interval), and by \Cref{obs:max-F-grid-interval}, each is non-empty.
Note that for each maximal $ \F_\rho $ grid interval $[g_1,g_2]$, we have that $g_1-\Delta$ and $g_2 + \Delta$ are endpoints of maximal $\bF_\rho$ grid interval. Therefore, $a,b \in J_1$
for each interval $ [a,b] \in \calS $, and by the setting of $\sigma$ and \Cref{claim:max-bF-zero},
$ \sigma(\Gamma_k(a)),  \sigma(\Gamma_k(b)) \in \bF_\rho $.

For each $[a,b] \in \calS $ and the corresponding $[g_1,g_2]$, let
$\alpha(a,b) = \E_{t_1}(g_1)$, $\beta(a,b) = \E_{t_1}(g_2)$, $\gamma(a,b) = \parity(t_1)$, $\gamma'(a,b)= \parity(t_1-\Delta)$.
We shall apply \Cref{condition:non-localA} and \Cref{condition:non-localB} to modify $\sigma$ on all
$[a,b] \in \calS $ ``in parallel'' as described next,
and set $\E'_0$ to be the resulting configuration.

For each $[a,b] \in \calS $
we first apply \Cref{condition:non-localB} with $z=a$, $\nu = \alpha(a,b)$, $\gamma =  \gamma(a,b)$
and $\gamma' = \gamma'(a,b)$. We  let $a'=z'$ (recall that $z' \in [z+1,z+2k+1] $
and $\tsigma(\Gamma_k(z')) \in \F_\rho$). Next we apply \Cref{condition:non-localB} in its second (symmetric) variant with $z=b$, $\nu = \beta(a,b)$, $\gamma =  \gamma(a,b)$
and $\gamma' = \gamma'(a,b)$. We let $b' = z'$ (recall that in this variant, $z' \in [z-2k-2,z-1]$, and here too $\tsigma(\Gamma_k(z')) \in \F_\rho$).
Finally we apply \Cref{condition:non-localA} on the modified configuration with $x=a'$ and $y=b'$.

\begin{figure}
	\centerline{\mbox{\includegraphics[width=1.05\textwidth]{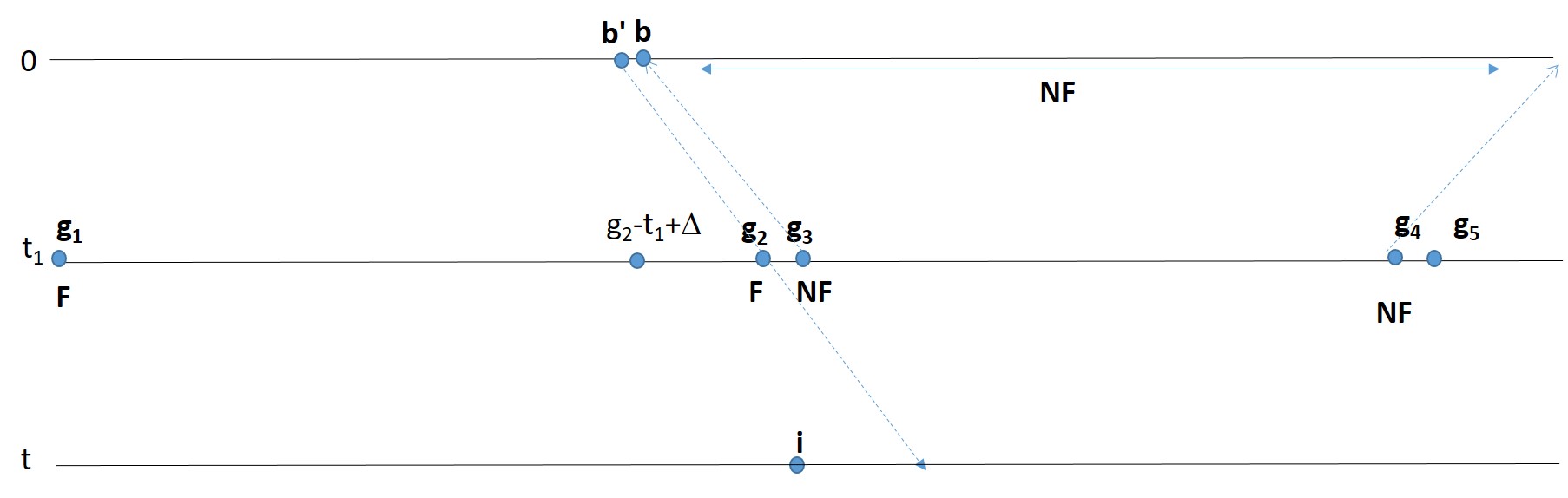}
	}}
	\caption{\small An illustration for the setting of $\E_0$.
		As in \Cref{fig:ABCU}, $[g_1,g_2]$ is a maximal $\F_\rho$ grid interval, $g_3 = g_2+\Delta$, where $[g_3,g_4]$ is a maximal $\bF_\rho$ grid interval, and $g_5=g_4+\Delta$ is an endpoint of a maximal $\F_\rho$ grid-interval.
		The maximal $\bF_\rho$ grid interval $[g_3,g_4]$ is used to set the locations between $g_3-t_1 = g_2-t_1+\Delta$ and $g_4+t_1$ (more precisely, between $g_3-t_1-k$ and $g_4+t_1+k$ based on \Cref{claim:max-bF-zero}. The location $b=g_2-t_1+\Delta$ is an endpoint of an interval in $\calS$, and the location $b'$ is determined by the application of
		\Cref{condition:non-localB}. The values in the $k$-neighborhood of $b'$ are set so that the evolution of $\rho$ will result in the $\E_{t_1}(g_2)$ at time $t_1$. The pair $(t,i)$ belongs to the set $B$.
	}
	\label{fig:E0}
\end{figure}

\subsection{The distance between $\E$ and $\E'$}\label{subsubsec:E-E-prime}
In this subsection we show that based on the counter-assumption regarding the number of violating pairs,
the number of pairs $(t,i) \in \cycnums{n}\times\nums{m}$ on which $\E$ and $\E'$ differ is at most $\eps m n$. To this end we show that each $(t,i)$ such that $\E_t(i) \neq \E'_t(i)$ belongs to one of the following sets:

\begin{enumerate}
	\item The set of pairs $(t,i)$ for which $ 0 \le t \le t_2 $.
	\item The uncertainty set $ U $.
	\item The set of $ (t,i) $ pairs where $ (t,i) $ is a violation with respect to $ \E_{t_1}[G] $.
\end{enumerate}
By the setting of $t_2$ ($t_1$) and $\Delta$, the number of pairs in the first set is at most
$\frac{(b_1+1)\eps}{b_0}  m n$.
By \Cref{claim:U_is_small}, $ |U| \leq \frac{5\eps}{b_1} m n $.
By our counter-assumption, the number of violating pairs is at most $\frac{\eps}{b_2}mn$ .
Setting $b_1 = 15$, $b_0 = 48$ and $b_2=3$, we get a total of at most $\eps m n$ pairs, as claimed.

To establish the claim that each $(t,i)$ for which $\E_t(i) \neq \E'_t(i)$ belongs to one of the above three sets, we
prove the contrapositive.
Suppose the pair $ (t,i) $ is not in the uncertainty set $ U $ and that $ t > t_2 $.
It follows that  $ (t,i) \in A \cup B \cup C $.
We show that for each of the three types of pairs 
($ (t,i) \in A $, $ (t,i) \in B $, and $ (t,i) \in C $), if the pair $ (t,i) $ is not a violating pair with respect to $ \E_{t_1}[G] $, it must hold that $ \E_t(i)=\E'_t(i) $.

\subparagraph{Pairs {\boldmath{$ (t,i) \in A $}}.}
By the definition of $A$, there exists a maximal $ \F_\rho $ grid interval $ [g_1(i), g_2(i)] $ (with respect to $ \E_{t_1} $) for which
$i \in [g_1(i)+ t_1,g_2(i)-t_1]$.
Since $ (t,i) $ is not a violating pair with respect to $ \E_{t_1}[G] $, it must hold that $ \E_{t_2}(\Gamma_k(i)), \E_t(\Gamma_k(i)) \in \F_\rho $ and that $ \E_t(i)= \frhor(\E_{t_2}(i), \parity(t-t_2), 0)  $.
Since 
$i\in [g_1(i)+t_1 , g_2(i)-t_1 ]$,
we know  that $ i \notin J $.
Hence, by the definition of the configuration $ \sigma $, we have that $ \sigma(i) = \frhor(\E_{t_2}(i), \parity(t_2), 0) $ and that $ \sigma(\Gamma_k(i)) \in \F_\rho $.
Let $[a(i),b(i)] = [g_1(i) - \Delta + t_1, g_2(i) + \Delta - t_1]$, so that $i \in I(a(i),b(i))$. By
the definition of $E'_0$, based on \Cref{condition:non-localA}
we have that $\E'_0(i) = \sigma(i)$ and $\E'_0(\Gamma_k(i)) \in \F_\rho$.
Since $\E'$ evolves according to $\rho$, by \Cref{condition:final_prediction},

\ifnum\fullversion=0
\vspace{-3.5ex}
\fi
\begin{align*}
	\E'_{t_2}(i) = \frhor(\E'_{0}(i), \parity(t_2), 0)
	= \frhor(\frhor(\E_{t_2}(i), \parity(t_2), 0), \parity(t_2), 0)
	= \E_{t_2}(i)
\end{align*}
where the last equality follows 
from (the second part of) \Cref{obs:f-lar}.
Additionally, by \Cref{condition:final}, $ \E'_{t_2}(i) \in \F_\rho $.
Therefore, by \Cref{condition:final_prediction},

\ifnum\fullversion=0
\vspace{-3.5ex}
\fi
\begin{align*}
	\E'_{t}(i) = \frhor(\E'_{t_2}(i), \parity(t-t_2), 0) = \frhor(\E_{t_2}(i), \parity(t-t_2), 0)  = \E_{t}(i)\;.
\end{align*}

\subparagraph{Pairs {\boldmath{$ (t,i) \in B $}.}}
By the definition of $B$, 
there exists a maximal $\F_\rho$ grid interval $ [g_1(i), g_2(i)] $ with respect to $\E_{t_1}$ such that either
$i \in [g_1(i) -(t-t_1),g_1(i)+t_1-\Delta-1]$ or $i \in [g_2(i)-t_1+\Delta+1,g_2(i)+(t-t_1)]$.
Assume (without loss of generality) that 
the latter holds.
By the definition of $B$ we also know that for every other  maximal $\F_\rho$ grid interval $ [g'_1, g'_2] $ 
it holds that $\dist(i,g_2(i)) < \dist(i,g_1'),\dist(i,g_2')- \Delta$. 
Let 
$g(i)$     
be the grid location closest to $i$ in $[g_2(i)-t_1+\Delta,g_2(i)]$.
Since 
$i \in [g_2(i)-t_1+\Delta+1,g_2(i)+(t-t_1)]$,
necessarily, $g(i) \in [g_2(i)-t_1,g_2(i)]$.
For the sake of conciseness, 
in what follows we shall use $g_1$, $g_2$, and $g$ as a shorthand for $g_1(i)$, $g_2(i)$ and $g(i)$, respectively.

Since $ [g_1, g_2] $ is a maximal $ \F_\rho $ grid interval, 
$ [a =g_1 - \Delta + t_1, b=g_2 + \Delta - t_1] \in \mathcal{S} $. Hence, to obtain $ \E'_0 $ from the configuration $ \sigma $, we  
invoked \Cref{condition:non-localB} (the symmetric version) with $z=b$, $\nu = \E_{t_1}(g_2)$,
$\gamma=\parity(t_1)$, and $\gamma' = \parity(t_1-\Delta) = \parity(\dist(g_2,b))$.
By \Cref{condition:non-localB}, letting $b' = z'$,
$ \E_{t_1}(g_2) = \frhor(\E'_0(b'), \parity(t_1), \parity(\dist(g_2,b'))) $.
By \Cref{obs:f-lar},

\ifnum\fullversion=0
\vspace{-3.5ex}
\fi
\begin{equation}\label{eq:E0bprime}
	\E'_0(b') = \frhol(\E_{t_1}(g_2), \parity(t_1), \parity(\dist(g_2,b')))\;.
\end{equation}

As $b' \in [b-2k-1,b-1]$, which by the setting of $b$ implies that $b' \in [g_2+\Delta-t_1-2k-1, g_2+\Delta-t_1-1]$,
we have that $(0,b')$ is an ancestor of $(t_1,j)$ for every $j \in [g_2-t_1,g_2]$.
In particular this holds for the grid location $g$ (that is closest to $i$ in $G\cap [g_2-t_1,g_2]$).
Since $(t,i)$ descends from $(t_1,g)$, we get that $(t,i)$ also descends from $(0,b')$.

We claim that for every $ b'' \ne b' $ with $ \dist(i,b'') < \dist(i,b') $ it holds that $ \E'_0(b'') \in \bF_\rho $.
To verify this, let $[g_3,g_4]$ be the maximal $\bF_\rho$ grid interval where $g_3 = g_2+\Delta$, and let $g_5 = g_4+\Delta$, so that $g_5$ is the endpoint of a maximal $\F_\rho$ grid interval. By the definition of $\E'_0$ (based on $\sigma$ and \Cref{condition:non-localB}) we have that $\E'_0(\Gamma_k(j)) \in \bF_\rho$ for every $j \in [b'+1,b-1] \cup J_1(g_3,g_4) =   [b'+1,g_4+t_1]$.
Since (by the second requirements on pairs in $B$), $\dist(i,g_2) < \dist(i,g_5) - \Delta$ 
and $b' \in [g_2-t_1 +\Delta-2k-1, g_2-t_1+\Delta-1]$,
we have that $ \E'_0(b'') \in \bF_\rho $ for every $ b'' \ne b' $ with $ \dist(i,b'') < \dist(i,b') $.
Therefore, we can apply \Cref{condition:final_prediction} to obtain that

\ifnum\fullversion=0
\vspace{-3.5ex}
\fi
\begin{align}
	&\E'_t(i) =\frhor(\E_0(b'), \parity(t), \parity(\dist(i,b'))) \nonumber \\
	&= \frhor(\frhor(\E_{t_1}(g_2), \parity(t_1), \parity(\dist(g_2,b'))), \parity(t), \parity(\dist(i,b'))) \label{eq:ig2a}\\
	&= \frhor(\E_{t_1}(g_2), \parity(t-t_1), \parity(\dist(i,g_2))) \label{eq:ig2b}
\end{align}
where the last equality follows from \Cref{obs:f-lar}.

Consider first the case that $g = g_2$.
Since the pair $ (t,i) $ is not a violating pair,

\ifnum\fullversion=0
\vspace{-3.5ex}
\fi
\begin{equation}\label{eq:ig2c}
	\E_t(i) \;=\; \frhor(\E_{t_1}(g_2), \parity(t-t_1), \parity(\dist(i,g_2)))  \;,
\end{equation}
and hence in this case,  $\E_t(i) = \E'_t(i)$, as desired.
We next turn to the case that $g \neq g_2$.
We claim that since $\E_{t_1}[G]$ is feasible,

\ifnum\fullversion=0
\vspace{-3.5ex}
\fi
\begin{equation}\label{eq:gbprime}
	\E_{t_1}(g) = \frhor(\E'_0(b'), \parity(t_1), \parity(\dist(g,b'))) \;.
\end{equation}
Conditioned on Equality~\ref{eq:gbprime} holding, the argument is the same as for the case that $g=g_2$ (replacing $g_2$ with $g$ in Equations~\eqref{eq:ig2a}--\eqref{eq:ig2c}).

To verify Equation~\eqref{eq:gbprime},
we introduce the notion of a \emph{source} for a final pair.
Let $\E''$ be an environment that evolves according to $\rho$, and $(t',i')$ a final pair with respect to
$\E''$ and $\rho$. We say that $(0,b'')$ is the \emph{source} of $(t',i')$ (at time $0$) in $\E''$ if  $(0,b'')$ is an ancestor of $(t',i')$,
is final, and $\dist(b'',i') < \dist(j,i')$ for every other final $(0,j)$.
Consider any feasible extension $E''$ of $\E_{t_1}[G]$. By \Cref{claim:max-bF-zero} and the discussion above, the source $(0,b'')$ of $(t_1,g_2)$ (at time $0$ in $\E''$) must satisfy $b'' \in [g_2-t_1,g_2-t_1+\Delta-1]$. Furthermore, $(0,b'')$ must also be the source of $(t_1,g')$ for every grid location $g' \in [g_2-t_1,g_2]$.
Therefore, for each such grid location,
$\E_{t_1}(g') = \E''_{t_1}(g') = \frhor(\E''_0(b''), \parity(t_1), \parity(\dist(g',b'')))$, where
$\E''_0(b'') = \frhol(\E_{t_1}(g_2), \parity(t_1), \parity(\dist(g_2,b'')))$.
If $\frhor$ and $\frhol$ do no depend on their third argument, then, by Equation~\eqref{eq:E0bprime},
$\E'_0(b') = \E''_0(b'')$ and  if they do, then
$\E'_0(b') = \E''_0(b'') \oplus \parity(\dist(b',b''))$ .
In either case, Equation~\eqref{eq:gbprime} follows.

\subparagraph{Pairs {\boldmath{$ (t,i) \in C $}}.}
By the definition of $C$, there exists a maximal 
$\bF_\rho$ grid interval $ [g_1(i), g_2(i)] $ 
such that  $ g_1(i) \leq i \leq g_2(i) $.
Additionally, the pair $ (t,i) $ does not descend from either the pair
$(t_1,g_1(i)-1) $ or from the pair $ (t_1,g_2(i)+1) $.
Let $ g(i)$ be the grid location defined in \Cref{def:C-violating} (of violating pairs in $C$), so that
$g(i) \in G \cap [g_1(i),g_2(i)]$ and
$\dist(i,g(i)) < \Delta$.

By the definition of $\E'_0$ (based on $\sigma$ -- recall Equation~\eqref{eq:sigma-J}), we have that
$ \E'_0(\Gamma_k(i)) = \hrhol(\E_{t_1}(\Gamma_k(g(i))),\parity(t_1), \ddist(i,g(i))) $.
By the definition of $\hrhol$ (\Cref{def:h-lar}), this implies that
$ \E_{t_1}(\Gamma_k(g(i))) = \hrhor(\E'_{0}(\Gamma_k(i)),\parity(t_1), \ddist(i,g(i))) $.
Since $\E'_0(\Gamma_k(j)) \in \bF_\rho$ for every location $j\in J(g_1(i)),g_2(i))$,
and the environment $\E'$ evolves according to $\rho$ where $\rho$ satisfies \Cref{condition:noninf_prediction},
we  know that
$ \E'_{t_1}(\Gamma_k(g(i))) = \hrhor(\E'_{0}(\Gamma_k(i)),\parity(t_1), \ddist(i,g(i))) $.
Hence, $\E'_{t_1}(\Gamma_k(g(i)))  = \E_{t_1}(\Gamma_k(g(i)))$.
Furthermore, using in addition the fact that $ (t,i) $ does not descend from either
$(t_1,g_1(i)-1) $ or $ (t_1,g_2(i)+1) $, we get all ancestors of $(t_1,j)$ of $(t,i)$
satisfy $\E'_{t_1}(\Gamma_k(j)) \in \bF_\rho$, so that
$ \E'_{t}(\Gamma_k(i)) = \hrhor(\E'_{t_1}(\Gamma_k(g(i))),\parity(t-t_1), \ddist(g(i),i))) $.
Since $(t,i)$ is not a violating pair,
$ \E_t(\Gamma_k(i)) = \hrhor(\E_{t_1}(\Gamma_k(g(i))), \parity(t-t_1), \ddist(g(i),i)) $,
and using $\E'_{t_1}(\Gamma_k(g(i)))  = \E_{t_1}(\Gamma_k(g(i)))$ we get that $\E'_{t}(i) = \E_t(i)$.

\ifnum\fullversion=1
\subsection{The homogeneous cases}\label{subsubsec:homogeneous}
Consider first the case that $\E_{t_1}(\Gamma_k(g))\in \bF_\rho$ for every grid location $g\in G$, which we refer to as the \emph{fully non-final} case.
In this case we have  the following variant of \Cref{claim:max-bF-zero} (whose proof is essentially the same as the proof of \Cref{claim:max-bF-zero}).
\begin{claim}\label{claim:max-bF-zero-homogeneous}
	Let $\rho$ be any local rule that satisfies Conditions~\ref{condition:final}--\ref{condition:non-localB}.
	Let $\E''$ be any environment that is a feasible extension of $\E_{t_1}[G]$ with respect to $\rho$, where
	$\E_{t_1}(\Gamma_k(g))\in \bF_\rho$ for every $g\in G$.
	Then $\E''_0(\Gamma_k(i)) \in \bF_\rho$ for every 
	$i \in \cycnums{n}$.
	Furthermore,  $\E''_0(\Gamma_k(i)) = \hrhol(\E_{t_1}(\Gamma_k(g)), \parity(t_1), \ddist(i,g))$
	for any $g \in G$ and every ancestor $(0,i)$ of $(t_1,g)$. 
\end{claim}
Therefore, in the fully non-final case we set $\E'_0$ as indicated by \Cref{claim:max-bF-zero-homogeneous} (and let $E'$ be the environment that evolves according to $\rho$ from $\E'_0$).
Recall that in this case, $C = \{(t,i): t_2 < t < m,\; i \in \cycnums{n}\}$ and the sets $A$ and $B$ (as well as $U$) are empty. Therefore, the only violating pairs $(t,i)$ are those that belong to $C$, and it suffices to show that $E'$ and $E$ only differ on these pairs, on top of the pairs $(t,i) \in \nums{t_2} \times \cycnums{n}$.
This is established as done in the more complex heterogeneous case (for $(t,i)\in C$), and recalled next.

Consider any pair $(t,i) \in C$, and
let $ g(i)$ be the grid location defined in \Cref{def:C-violating} (of violating pairs in $C$).
so that
$\dist(i,g(i)) < \Delta$.
By the definition of $\E'_0$ (based on $\sigma$ -- recall Equation~\eqref{eq:sigma-J}), we have that
\[ \E'_0(\Gamma_k(i)) = \hrhol(\E_{t_1}(\Gamma_k(g(i))),\parity(t_1), \ddist(i,g(i))) \;.\]
By the definition of $\hrhol$ (\Cref{def:h-lar}), this implies that
\[ \E_{t_1}(\Gamma_k(g(i))) = \hrhor(\E'_{0}(\Gamma_k(i)),\parity(t_1), \ddist(i,g(i))) \;. \]
Since $\E'_0(\Gamma_k(j)) \in \bF_\rho$ for every location $j \in \cycnums{n}$,
and the environment $\E'$ evolves according to $\rho$ where $\rho$ satisfies Condition~\ref{condition:noninf_prediction},
we  know that
\[ \E'_{t_1}(\Gamma_k(g(i))) = \hrhor(\E'_{0}(\Gamma_k(i)),\parity(t_1), \ddist(i,g(i))) \;. \]
Hence, $\E'_{t_1}(\Gamma_k(g(i)))  = \E_{t_1}(\Gamma_k(g(i)))$.
Furthermore,
we have that all ancestors of $(t_1,j)$ of $(t,i)$
satisfy $\E'_{t_1}(\Gamma_k(j)) \in \bF_\rho$, so that
\[ \E'_{t}(\Gamma_k(i)) = \hrhor(\E'_{t_1}(\Gamma_k(g(i))),\parity(t-t_1), \ddist(g(i),i))) \;. \]
Since $(t,i)$ is not a violating pair,
\[ \E_t(\Gamma_k(i)) = \hrhor(\E_{t_1}(\Gamma_k(g(i))), \parity(t-t_1), \ddist(g(i),i))  \;, \]
and using $\E'_{t_1}(\Gamma_k(g(i)))  = \E_{t_1}(\Gamma_k(g(i)))$ we get that
$\E'_{t}(i) = \E_t(i)$, as desired.

\medskip
We now turn to the case that $\E_{t_1}(\Gamma_k(g))\in \F_\rho$ for every $g\in G$, which we refer to as the \emph{fully final} case. In this case we initialize $\sigma(i) = \frhor(\E_{t_2}(i), \parity(t_2), 0)$ for each $i\in \cycnums{n}$, and apply \Cref{condition:non-localA} (on the complete interval ranging from $a=0$ and back to $b=0$).
Recall that in this fully final case, $A  = \{(t,i): t_2 < t < m,\; i \in \cycnums{n}\}$ and the sets $B$ and $C$ (as well as $U$) are empty. Therefore, the only violating pairs $(t,i)$ are those that belong to $A$, and it suffices to show that $E'$ and $E$ only differ on these pairs (on top of the pairs $(t,i) \in \nums{t_2} \times \cycnums{n}$).
This is established as done in the more complex heterogeneous case (for $(t,i)\in A$), and recalled below.

Consider any pair $(t,i)\in A$ that is non-violating.
Since $ (t,i) $ is not a violating pair with respect to $ \E_{t_1}[G] $, it must hold that $ \E_{t_2}(\Gamma_k(i)), \E_t(\Gamma_k(i)) \in \F_\rho $ and that $ \E_t(i)= \frhor(\E_{t_2}(i), \parity(t-t_2), 0)  $.
By the definition of the configuration $ \sigma $, we have that $ \sigma(i) = \frhor(\E_{t_2}(i), \parity(t_2), 0) $ and that $ \sigma(\Gamma_k(i)) \in \F_\rho $.
By \Cref{condition:non-localB} and the definition of $\E'_0$ we have that $\E'_0(i) = \sigma(i)$ and $\E'_0(\Gamma_k(i)) \in \F_\rho$.
Since $\E'$ evolves according to $\rho$, by \Cref{condition:final_prediction},
\[\E'_{t_2}(i) = \frhor(\E'_{0}(i), \parity(t_2), 0) =
\frhor(\frhor(\E_{t_2}(i), \parity(t_2), 0), \parity(t_2), 0) = \E_{t_2}(i)
\]
where the last equality follows
from (the second part of) \Cref{obs:f-lar}.
Additionally, by \Cref{condition:final}, $ \E'_{t_2}(i) \in \F_\rho $.
Therefore, by \Cref{condition:final_prediction},
\[
\E'_{t}(i) = \frhor(\E'_{t_2}(i), \parity(t-t_2), 0) = \frhor(\E_{t_2}(i), \parity(t-t_2), 0)  = \E_{t}(i)\;,
\]
as required.
\fi

\ifnum\fullversion=1

\input{smaller-m}

\input{rules}

\fi

\ifnum\fullversion=1
\bibliography{ref}
\else
\bibliography{ref-short}
\fi

\end{document}

%% file: includes.tex
\usepackage{nameref}
\usepackage{hyperref}
\hypersetup{colorlinks={true},linkcolor={blue},citecolor=blue}

%% file: defines.tex
\numberwithin{equation}{section}
\numberwithin{figure}{section}
\newtheorem{condition}{Condition}
\newtheorem{openproblem}{{\sf Open Problem}}

\newcommand{\set}[1]{\left\{ #1 \right\}}

\newcommand{\NN}{\mathbb{N}}

\newcommand{\calS}{\mathcal{S}}

\newcommand{\dana}[1]{\ifnotes {\noindent \scriptsize  \textcolor{blue} {Dana: {#1}}} \fi{}}
\newcommand{\parity}{\textsf{parity}}
\newcommand{\maj}{{\textrm{MAJ}}}
\newcommand{\orr}{{\textrm{OR}}}

\newcommand{\norr}{{\textrm{NOR}}}

\newcommand{\mino}{{\textrm{MIN}}}
\newcommand{\fuh}{{\textrm{FUH}}}
\newcommand{\fih}{{\textrm{FIH}}}

\newcommand{\bitset}{\{0,1\}}
\newcommand{\eps}{{\epsilon}}
\newcommand{\poly}{\text{poly}}
\newcommand{\nums}[1]{[#1]}
\newcommand{\cycnums}[1]{{\mathbb{Z}_{#1}}}
\newcommand{\E}{{\tt ENV}}
\newcommand{\calE}{\mathcal{E}}
\newcommand{\ddist}{\overrightarrow{\rm dist}}
\newcommand{\dist}{{\rm dist}}
\newcommand{\lar}{\leftarrow}

\newcommand{\frhor}{f_\rho}
\newcommand{\frhol}{f_\rho^{\lar}}
\newcommand{\hrhor}{h_\rho}
\newcommand{\hrhol}{h_\rho^{\lar}}
\newcommand{\fruler}[1]{f_{#1}}

\newcommand{\hruler}[1]{h_{#1}}

\newcommand{\F}{{\mathcal{F}}}

\newcommand{\bF}{{\overline{\cal F}}}
\newcommand{\M}{{\mathcal{M}}}
\newcommand{\tsigma}{{\widetilde{\sigma}}} 

%% file: smaller-m.tex
\section{The case of $m\ll n$}\label{subsec:m-n}

Let $\rho$ be a fixed local rule that satisfies Conditions~\ref{condition:final}--\ref{condition:non-localB}.

In order to address the case that $m$ is much smaller than $n$ (so that the multiplicative factor of $n/m$
in the complexity stated in \Cref{thm:main} is too large), we essentially reduce to the case that $m = \Theta(\eps n)$, as explained next. 
Let $n' = b_3 m/\eps$ (for a sufficiently large constant $b_3$), and assume that $n > b_3 m/\eps^2$, or else we run the  algorithm given in \Cref{subsec:alg}, so that we get query complexity $O(1/\eps^4)$.
Also, assume for simplicity that $n$ is divisible by $n'$. We partition $\cycnums{n}$ into consecutive disjoint intervals of size $n'$ each. The algorithm selects $\Theta(1/\eps)$ intervals, uniformly at random and for each selected interval it runs a test to check whether the interval evolves according to $\rho$. We next give more precise details about how such a test is performed. At this point, we just note that the test for each selected interval performs $O(1/\eps^3)$ queries, so that here too we get a total of $O(1/\eps^4)$ queries.

Consider any such interval of size $n'$, denoted $I= [i_0,j_0=i_0+n'-1]$. Here too, $\Delta = \eps^2 m/b_0$, where we assume that $n'$ is divisible by $\Delta$, $t_1 = \frac{b_1 \Delta}{\epsilon} $, and $t_2 = t_1 + \Delta$. Observe that $n'/\Delta = O(1/\eps^3)$, and that $t_2\cdot n' = O(\eps mn')$.
For $i'_0 = i_0+t_1+k+1$ and $j'_0 = j_0 - t_1-k-1$,
we let $G_I = \{i'_0,i'_0+\Delta,\dots,j'_0\}$
be the grid associated with $I$ (we assume for simplicity that $j_0-i_0$ is divisible by $\Delta$).
The algorithm starts by querying $\E_{t_1}$ on all locations in $\Gamma_k(G_I)$. If $ \E_{t_1}[G_I] $ is infeasible with respect to $\rho$, then the algorithm rejects. Otherwise, the algorithm selects, uniformly at random, $ \Theta(\frac{1}{\epsilon}) $ pairs $ (t,i) $ where $ t_2 < t < m $ and $ i \in \{i'_0+(t-t_1),\dots,j'_0-(t-t_1)\}$ (which equals $\{i_0+k+1+t,j_0-k-1-t\}$). For each selected pair $ (t,i) $, it queries $ \E_{t}(\Gamma_k(i)) $ and $ \E_{t_2}(\Gamma_k(i)) $. If some  selected pair is a violating pair with respect to $\rho$ (as defined in \Cref{subsec:violating_pairs}), then it rejects, and otherwise, it accepts.
Note is that for each interval $I=[i_0,j_0]$ and for every $ (t,i) $ where $ t_2 < t < m $ and $ i \in \{i'_0+(t-t_1),\dots,j'_0-(t-t_1)\}$, the pair $(t,i)$ cannot descend form a pair $(0,\ell)$ such that $\ell \notin I$.

 If $\E$ evolves according to $\rho$, then the algorithm never rejects, as $\E_{t_1}[G_I]$ is feasible for every $I$, and there are no violating pairs with respect to $\E_{t_1}[G_I]$ and $\rho$, for every $I$ (following the proof of \Cref{lemma:completeness}).
 It remains to verify that if the algorithm accepts $\E$ with probability at least $2/3$, then $\E$ is $\eps$-close to some $\E'$ that evolves according to $\rho$.

  We say that an interval $I= [i_0,j_0]$ is \emph{bad} if either $ \E_{t_1}[G_I] $ is infeasible with respect to $\rho$ or there are more than $ (\eps/b_4) m n' $ violating pairs $(t,i)$ (for $ t_2 < t < m $ and $ i \in \{i'_0+(t-t_1),\dots,j'_0-(t-t_1)\}$, where $i'_0$ and $j'_0$ are as defined above and $b_4$ is a sufficiently large constant). Otherwise it is \emph{good}. Since the algorithm accepts $\E$ with  probability at least $2/3$, the fraction of bad intervals is at most $\eps/b_5$ (for an appropriate constant $b_5$). For each good interval $[i_0,j_0]$ we set $\E'_0$ restricted to (a sub-interval of) $[i_0'-(t_1+k),j'_0-(t_1+k)]$ as in the proof of \Cref{lemma:soundness} (in \Cref{subsubsec:E-prime}). All remaining locations in $\E'_0$ (i.e., $[i_0,i'_0-(t_1+k)-1]$ and
  $[j'_0+(t_1+k)+1,j_0]$ in each good interval $I=[i_0,j_0]$, and all of $I$ for each bad interval $I'$), are set arbitrarily.
  We let $\E'$ be the environment that evolves from $\E'_0$ (according to $\rho$).

  Following the second part of the proof of \Cref{lemma:soundness} (\Cref{subsubsec:E-E-prime}), for each good interval $I = [i_0,j_0]$, the environments $\E$ and $\E'$ differ on at most all locations in $\nums{t_2}\times I$
  (whose number is $t_2\cdot n' \leq \frac{(b_1+1) \Delta}{\epsilon} \cdot n' = \frac{b_1+1}{b_0}\cdot \eps mn'$), on at most all locations in $U$ restricted to $I$ (whose number is at most $\frac{5}{b_1}\eps mn' $), and on at most all violating pairs restricted to $I$ (whose number is at most $\frac{1}{b_4}\eps mn'$). In addition, they differ on at most all pairs $(t,i)$ where $t_2+1 \leq t < m$ and $i \in [i_0,i_0+2m] \cup [j_0-2m,j_0]$, whose number is upper bounded by $4m^2 = \frac{4}{b_3}\cdot \eps mn'$.
  They agree on all other locations in $\nums{m}\times I$. For each bad interval $I'$ they may disagree on all locations in $\nums{m}\times I$. 
  Summing up all disagreements (setting e.g., $b_1=20$, $b_0 = 84$, $b_4=4$, $b_3=32$, and $b_5=8$), we get at most $\eps m n$, as claimed.

%% file: rules.tex

\section{Threshold rules}\label{subsec:thresh-rules}
\paragraph*{Complementary Rules.}
Some rules are equivalent to each other in the sense that if we interchange the roles of 0 and 1 in one rule, we get the other.
Formally, two rules $ \rho, \rho':\bitset^3 \to \bitset $ are said to be \emph{complementary} if for every triplet $ (\beta_1,\beta_2,\beta_3) \in \bitset^3 $, it holds that $ \rho(\beta_1,\beta_2,\beta_3) = 1-\rho'(1-\beta_1,1-\beta_2,1-\beta_3) $. Clearly, complementary rules are equivalent for testing purposes.
Observe that some rules are the complement of themselves, so the \textit{complement} is an equivalence relation that partitions the rules into pairs and singletons.

\paragraph*{Threshold Rules.}
Recall that a rule $ \rho :\bitset^3 \to \bitset $ is a \emph{threshold} rule if there exist a threshold integer
$ 0 \leq b \leq 3 $ and a bit $ \alpha \in \set{0, 1} $ such that $ \rho(\beta_1,\beta_2,\beta_3) = \alpha $ if and only if $ \beta_1+\beta_2+\beta_3 \geq b $.

\medskip\noindent
Since the all-$1$ and all-$0$ rules are complementary, and similarly OR and AND as well as NOR and NAND, it suffices to consider the rules: all-$1$, OR, NOR, Majority and Minority.
The all-$1$  rule can clearly be tested with $O(1/\eps)$ queries (selected uniformly in $[m-1]\times \cycnums{n}$), as it converges in a single time step to the all-$1$ configuration.
Turning to NOR, we show that it converges in a single time step, from which an $O(1/\eps)$-query algorithm easily follows.
For the remaining, non-trivial, rules we show that they satisfy Conditions~\ref{condition:final}--\ref{condition:non-localB}.

\subsection{The NOR and NAND rules}\label{subsec:NOR}
We show that the NOR rule converges after at most a single time step.
Since the NAND rule is equivalent, an analogous proof holds for the NAND rule as well.

We first claim that if at some time step $ t $, every block of consecutive $ 0 $'s in the configuration $ \E_t $ is of size at least 3, then $ \E_{t+2} = \E_{t} $.
Let $ i \in \cycnums{n} $ be a location.
We show that $ \E_{t+2}(i) = \E_{t}(i) $.
If $ \E_t(i)=1 $, then necessarily $ \E_{t+1}(i) = \E_{t+1}(i-1) = \E_{t+1}(i+1) = 0 $, and therefore $ \E_{t+2}(i) = \norr(0,0,0) = 1 = \E_t(i) $.
Next consider the case that $ \E_t(i)=0 $.
If $ \E_t(i-1)=\E_t(i+1)=0 $, then $ \E_{t+1}(i)=1 $, and then $ \E_{t+2}(i)=0 $ as required.
Suppose this is not the case.
That is, the value of either $ \E_t(i-1) $ or $ \E_t(i+1) $ is $ 1 $.
Without loss of generality, assume $ \E_t(i-1)=1 $.
In this case, it must hold that $ \E_t(i+1) = \E_t(i+2)=0 $, or else the location $ i $ would have been part of a $ 0 $-block of size less than 3.
Now, $ \E_{t+1}(i+1) = \norr(0,0,0) = 1 $, and therefore $ \E_{t+2}(i)=0 $.

Now we claim that if $ t>0  $, there are no $ 0 $-blocks of length less than 3 in $ \E_t $.
Suppose by way of contradiction that for some $ t>0 $, the configuration $ \E_t $ contains the pattern $ 101 $ or $ 1001 $.
Let $ i \in \cycnums{n} $ be the left-most location and $ j \in \cycnums{n} $ be the right-most location in the ``bad'' pattern.
Since $ \E_t(i)=\E_t(j)=1 $, it must hold that $ \E_{t-1}(i-1) = \E_{t-1}(i) = \E_{t-1}(i+1) = 0 $ and that $ \E_{t-1}(j-1) = \E_{t-1}(j) = \E_{t-1}(j+1) = 0 $.
In the case of the pattern being $ 101 $, $ i+1=j-1 $.
Hence, in this case, $ \E_t(i+1) = \norr(\E_{t-1}(i), \E_{t-1}(i+1), \E_{t-1}(j)) = \norr(0,0,0) = 1 $.
In the case of the pattern being $ 1001 $, $ i+2=j-1 $.
Hence, in this case, $ \E_t(i+1) = \norr(\E_{t-1}(i), \E_{t-1}(i+1), \E_{t-1}(j-1)) = \norr(0,0,0) = 1 $.
That is, in both cases, $ \E_t(i+1) = 1 $ in contradiction to our conclusion that $ \E_t(i+1) = 0 $.

Now, since a $ 0 $-block of length less than 3 can only appear in the initial configuration, and cannot arise in any other way, and, as we've shown, once there are no 0-blocks of length less than 3, the environment alternates between a pair of configurations (that is, it has converged), any environment that evolves according to the NOR rule converges after at most a single time step, and hence is trivially $ \poly(1/\eps) $-testable.

\subsection{The OR and AND rules}\label{subsec:OR}
We prove that all conditions hold for the rule $\orr(\beta_1,\beta_2,\beta_3)= \beta_1\vee \beta_2 \vee \beta_3$.
Since the AND rule is equivalent to the OR rule, all the conditions must hold for the AND rule as well.

For $\orr$, $k=0$, $\F_{\orr}  = \{1\}$ (and $\bF_{\orr} = \{0\}$). This rule ultimately converges to the all-$1$ configuration, unless the starting configuration is the all-$0$ configuration (in which case it remains in this configuration indefinitely).
Since $k=0$, rather than writing $\E_t(\Gamma_0(i))$, we simply write $\E_t(i)$.
Turning to the conditions (where the numbering of the items below correspond to the numbering of the conditions).
\begin{enumerate}
\item If $\E_t(i) \in \F_{\orr}$, then $\E_t(i) =1$, so that $\E_{t+1}(i)=1$ as well (for any setting of $\E_t(i-1)$ and $\E_{t-1}(i+1)$).

\item If $\E_t(i) \in \bF_{\orr}$, then $\E_t(i) = 0$. If either $\E_t(i-1) = 1$ or $\E_t(i+1) = 1$ (or both), then $\E_{t+1}(i) = 1 \in \F_{\orr}$, and otherwise $\E_{t+1}(i) =0 \in \bF_{\orr}$.

\item The function $\fruler{\orr}$ simply equals its first argument (which is $1$ whenever the function is applied).

\item The function $\hruler{\orr}$ simply equals its first argument (which is necessarily $0$).

\item Let $ \sigma : \cycnums{n} \to \bitset $ be a configuration and let $ [x,y] $ be an interval
of locations such that $ \sigma(x) \in  \F_\orr $ and $ \sigma(y) \in \F_\orr $.
That is, $ \sigma(x) = \sigma(y) = 1 $.
We simply set $ \tsigma(i) = 1 $ for every $ i \in [x,y] $ and $ \tsigma(i) = \sigma(i) $ otherwise.
It clearly holds that for every $i \in [x,y]$ we have that $ \tsigma(i) = 1 \in \F_\orr $, and if $i \in [x,y]$ and $\sigma(i) \in \F_\orr$, then $ \tsigma(i) = 1 = \sigma(i) $, and so the requirements of \Cref{condition:non-localA} hold.

\item By the premise of \Cref{condition:non-localB} (regarding $z$), we have that $\sigma(z) = 0$.
We set $z' = z+1$ (in the symmetric version, $z' = z-1$) and $\tsigma(z') = 1$.
\end{enumerate}

\subsection{The Majority rule}\label{subsec:maj}
Here we repeat what was stated in \Cref{subsec:conditions} regarding the majority rule, denoted $\maj$, and add missing explanations when needed.

For this rule, $k=1$, and $\F_{\maj}  = \{111,110,011,000,001,100\}$ (so that $\bF_{\maj} = \{101,010\}$). This rule ultimately converges to configurations that constitute of intervals of at least two consecutive $1$s and intervals of at least two consecutive $0$s, unless the starting configuration is $(01)^{n/2}$ (in which case it alternates between this configuration and the complementary one $(10)^{n/2}$).

We now verify that the conditions hold.
\begin{enumerate}
\item If $\E_t(\Gamma_1(i)) = 111$, then $\E_{t+1}(\Gamma_1(i)) = 111 \in \F_{\maj}$,
if $\E_t(\Gamma_1(i)) = 110$, then $\E_{t+1}(\Gamma_1(i)) \in \{110,111\} \subset \F_{\maj}$, and if $\E_t(\Gamma_1(i)) = 110$, then $\E_{t+1}(\Gamma_1(i)) \in \{110,111\} \subset \F_{\maj}$ (analogous statements hold for $\E_t(\Gamma_1(i)) \in \{000,001,100\}$).

\item  Consider first the case that $\E_t(\Gamma_1(i)) = 101$ (so that it belongs to $\bF_{\maj})$.
In this case,  $\E_t(\Gamma_1(i-1)) \in \{110,010\}$ and $\E_t(\Gamma_1(i+1))\in \{011,010\}$.
If $\E_t(\Gamma_1(i-1))=110$ (which belongs to $\F_{\maj}$), then $\E_t(i-1)=1$ and $\E_t(i)=1$, so that
$\E_{t+1}(\Gamma_1(i)) \in \{110,111\} \subset \F_{\maj}$ (and the case that $\E_t(\Gamma_1(i+1))=011$ is analogous).
On the other hand, if both $\E_t(\Gamma_1(i-1))=010$ and $\E_t(\Gamma_1(i+1))=010$ (so that they both belong to $\bF_{\maj}$), then
$\E_{t+1}(\Gamma_1(i)) = 010$ (and it belongs to $\bF_{\maj}$ as well).

\item For any $\beta \in \bitset$, we have that $\fruler{\maj}(\beta,\cdot,\cdot)=\beta$.
To verify this, observe that under the premise of the condition, one of the following holds.
(1) $\E_{t'}(\Gamma_1(i')) = \beta\beta \overline{\beta} $, and for every ancestor $(t',i'')\neq (t',i')$ of $(t,i)$ for which $ \dist(i,i'') \le \dist(i,i') $ it holds that $ \E_{t'}(i'') = \overline{\parity(i''-i)}$.
(2) $\E_{t'}(\Gamma_1(i')) = \overline{\beta}\beta\beta  $ and for every ancestor $(t',i'')$ of $(t,i)$ such $ \dist(i,i'') \le \dist(i,i') $ it holds that $ \E_{t'}(i'') = \overline{\parity(i''-i)}$.
To illustrate the first case, suppose that $t-t' = 5$, then $\E_{t'}$ between locations $i-5$ and $i+5$ may be of the form $01110101010$ (so that at time $t'+1$ locations $i-4$ to $i+4$ are $111101010$, and time $t'+2$ locations $i-3$ to $i+3$ are $1111010$, so that $(t'+2,i)$ has become final with respect to $\E'$ and $\rho$, and $\E_{t''}(i)=1$ for every $t'' \geq t'+2$.

\item The function $\hruler{\maj}$ is defined as follows:
$\hruler{\maj}(010,\beta,x)=010$ if $\beta\oplus \parity(x) = 0$ and $\hruler{\maj}(010,\beta,x)=101$ if
$\beta\oplus \parity(x) = 1$. Similarly, $\hruler{\maj}(101,\beta,x)=101$ if $\beta\oplus \parity(x) = 0$
and $\hruler{\maj}(101,\beta,x)=010$ if $\beta\oplus \parity(x) = 1$.

\item
Let $ \sigma : \cycnums{n} \to \bitset $ be a configuration and let $ [x,y] $ be an interval
of locations such that $ \sigma(\Gamma_1(x)) \in  \F_\maj $ and $ \sigma(\Gamma_1(y)) \in \F_\maj $.
We define the configuration $ \tsigma $ as follows.
Let $ i \in \nums(n) $.
If $ i \notin [x,y] $, we set $ \tsigma(i) = \sigma(i) $.
Otherwise, let $ j_i \in [x,y] $ be the location that minimizes $ \dist(i,j) $ among the locations for which $ \sigma(\Gamma_1(j)) \in \F_\maj $, and set $ \tsigma(i) = \sigma(j_i) $.
In particular, note that if $ \sigma(\Gamma_1(i)) \in \F_\maj $, then $ \tsigma(i) = \sigma(i) $.
We claim that for every $i \in [x,y]$ we have that $ \tsigma(\Gamma_1(i)) \in \F_\maj $, and that if $\sigma(\Gamma_1(i)) \in \F_\maj$, then $ \tsigma(i)=\sigma(i) $.
Let $ i \in [x,y] $.
If $ \sigma(\Gamma_1(i)) \in \F_\maj $, then $ j_i=i $, and hence either $ \sigma(\Gamma_1(i-1)) \in \F_\maj $ or $ \sigma(\Gamma_1(i+1)) \in \F_\maj $, which implies that either $ \tsigma(i)=\tsigma(i-1) $ or $ \tsigma(i)=\tsigma(i+1) $, and in both cases, it holds that $ \tsigma(\Gamma_1(i)) \in \F_\maj $ and $ \tsigma(i)=\sigma(i) $.
If $ \sigma(\Gamma_1(i)) \in \bF_\maj $, then, assume without loss of generality that $ j^* \in [i+1,y] $ (the case in which $ j^* \in [x,i-1] $ is similar).
In this case, $ j_i = j_{i+1} $, and hence $ \tsigma(i) = \tsigma(i+1) $, which implies that $ \tsigma(\Gamma_1(i)) \in \F_\maj $.

\item
 Assume first that $\sigma(\Gamma_1(z)) = 101$. If $\nu = 1$, then we set $z'=z+1$ and $\tsigma(z+2) = 1$,  and if $\nu = 0$, then we set $z' = z+2$, $\tsigma(z+2)=\tsigma(z+3) = 0$.
In the first case, $\tsigma(\Gamma_1(z'))=011 \in \F_{\maj}$ and in the second $\tsigma(\Gamma_1(z'))=100 \in \F_{\maj}$ and $\tsigma(\Gamma_1(z+1)) = 010 \in \bF_{\maj}$. In both cases,
$ \frhor(\tsigma(z'),\gamma,\parity(z'-z)\oplus \gamma') = \tsigma(z') = \nu $ for any $\gamma$ and $\gamma'$.
If $\sigma(\Gamma_1(z)) = 010$, then for $\nu = 0$ we set $z'=z+1$ and $\tsigma(z+2) = 0$,  and for $\nu = 1$, we set $z' = z+2$, $\tsigma(z+2)=\tsigma(z+3) = 1$.
The symmetric variant of this condition is established similarly.

\end{enumerate}

\subsection{The Minority rule}\label{subsec:min}
\sloppy
For the minority rule, denoted $\mino$,  like the majority rule, $k=1$, and $\F_{\mino} = \{111,110,011,000,001,100\}$ (so that $\bF_{\mino} = \{101,010\}$). Ultimate Convergence is also similar to the majority rule, except that in each time step, blocks of $1$s ``flip'' to become blocks of $0$s and vice versa, and if the initial configuration is $(01)^{n/2}$, then it does not change.

We now turn to the conditions.
\begin{enumerate}
\item If $\E_t(\Gamma_1(i)) = 111$, then $\E_{t+1}(\Gamma_1(i)) = 000 \in \F_{\mino}$,
if $\E_t(\Gamma_1(i)) = 110$, then $\E_{t+1}(\Gamma_1(i)) \in \{001,000\} \subset \F_{\mino}$, and if $\E_t(\Gamma_1(i)) = 011$, then $\E_{t+1}(\Gamma_1(i)) \in \{100,000\} \subset \F_{\mino}$ (analogous statements hold for $\E_t(\Gamma_1(i)) \in \{000,001,100\}$).

\item Consider first the case that $\E_t(\Gamma_1(i)) = 101$ (so that it belongs to $\bF_{\mino})$.
In this case,  $\E_t(\Gamma_1(i-1)) \in \{110,010\}$ and $\E_t(\Gamma_1(i+1))\in \{011,010\}$.
If $\E_t(\Gamma_1(i-1))=110$ (which belongs to $\F_{\mino}$), then $\E_t(i-1)=0$ and $\E_t(i)=0$, so that
$\E_{t+1}(\Gamma_1(i)) \in \{001,000\} \subset \F_{\mino}$ (and the case that $\E_t(\Gamma_1(i+1))=011$ is analogous).
On the other hand, if both $\E_t(\Gamma_1(i-1))=010$ and $\E_t(\Gamma_1(i+1))=010$ (so that they both belong to $\bF_{\mino}$), then
$\E_{t+1}(\Gamma_1(i)) = 101$ (which is the same as $\E_t(\Gamma_1(i))$, and it belongs to $\bF_{\mino}$ as well).

\item For any $\beta \in \bitset$, we have that $\fruler{\mino}(\beta_1,\beta_2,\cdot) = \beta_1 \oplus \beta_2$.
This can be verified similarly to what was shown for $\maj$.

\item The function $\hruler{\mino}$ is defined as follows:
$\hruler{\mino}(010,\beta,x)=101$ if $\beta\oplus \parity(x) = 0$ and $\hruler{\mino}(010,\beta,x)=010$ if
$\beta\oplus \parity(x) = 1$. Similarly, $\hruler{\mino}(101,\beta,x)=010$ if $\beta\oplus \parity(x) = 0$
and $\hruler{\mino}(101,\beta,x)=101$ if $\beta\oplus \parity(x) = 1$.

\item This item is the same as the corresponding one for $\maj$.

\item
This item is very similar to the corresponding one for $\maj$.
The only difference is that there is a dependence on the parameter $\gamma$.
Specifically, if $\gamma=0$, then the setting is exactly as for $\maj$.
Suppose $\gamma=1$.
We assume first that $\sigma(\Gamma_1(z)) = 101$.
If $\nu = 0$, then we set $z'=z+1$ and $\tsigma(z+2) = 1$,  and if $\nu = 1$, then we set $z' = z+2$, $\tsigma(z+2)=\tsigma(z+3) = 0$.
In the first case, $\tsigma(\Gamma_1(z'))=011 \in \F_{\mino}$ and in the second $\tsigma(\Gamma_1(z'))=100 \in \F_{\mino}$ and $\tsigma(\Gamma_1(z+1)) = 010 \in \bF_{\mino}$.
In both cases, $ \frhor(\tsigma(z'),\gamma,\parity(z'-z)\oplus \gamma') = \tsigma(z') \oplus \gamma = 1 - \tsigma(z') = \nu $ for any $\gamma'$.
If $\sigma(\Gamma_1(z)) = 010$, then for $\nu = 1$ we set $z'=z+1$ and $\tsigma(z+2) = 0$,  and for $\nu = 0$, we set $z' = z+2$, $\tsigma(z+2)=\tsigma(z+3) = 1$.

\end{enumerate}

\section{Other rules}\label{subsec:other-rules}

\subsection{Flip if homogeneous}

This rule, denoted $\fih$, is defined as follows: $\fih(\beta_1,\beta_2,\beta_3) = \beta_2$ unless
$\beta_1=\beta_2=\beta_3$, in which case $\fih(\beta_1,\beta_2,\beta_3) = \overline{\beta}_2$.
This rule ultimately converges to configurations with blocks of size one or two (unless the initial configuration is the all-$0$ configuration or the all-$1$ configuration, in which case it alternates between the two).
Here we have $\bF_{\fih} = \{000,111\}$ (so that $\F_{\fih} = \{01?,?10,10?,?01\}$.

\begin{enumerate}
\item  If $\E_t(\Gamma_1(i)) \in \bF_{\fih}$ then either $\E_t(i) \neq \E_t(i-1)$ or
$\E_t(i) \neq \E_t(i+1)$ (possibly both), so that in the first case  $\E_{t+1}(i-1) = \E_t(i-1)$ and in the second case, $\E_{t+1}(i) = \E_t(i)$. In both cases $\E_{t+1}(i+1) = \E_t(i+1)$, so that $\E_{t+1}(\Gamma_1(i)) \in \F_{\fih}$ as required.

\item If $\E_t(\Gamma_1(i)) \in \bF_{\fih}$, then $\E_t(\Gamma_1(i)) \in \{000,111\}$. If both $\E_t(\Gamma_1(i-1))\in \bF_{\fih}$ and $\E_t(\Gamma_1(i-1))\in \bF_{\fih}$, then $\E_t(\Gamma_2(i)) \in \{00000,11111\}$, so that $\E_{t+1}(\Gamma_1(i))\in \{111,000\} = \bF_{\fih}$. On the other hand, if $\E_t(\Gamma_1(i-1))\in \F_{\fih}$ or $\E_t(\Gamma_1(i+1)) \in \F_{\fih}$, then $\E_t(\Gamma_2(i)) \in \{1000?,?0001,0111?,?1110\}$ so that $\E_{t+1}\in \{01?,?10,10?,?10\} = \F_{\fih}$.

\item The function $\frhor$ is defined as follows: $\frhor(\beta_1,\beta_2,\beta_3) = \beta_1 \oplus \beta_3$.
To verify this consider any $(t,i)$ and $(t',i')$ that satisfy the requirements in \Cref{condition:final_prediction}. Assume that $\dist(i',i) = \ddist(i',i) $ (the case that $\dist(i',i) = \ddist(i,i')$ is verified similarly.
Since $\E_{t'}(\Gamma_1(i')) \in \F_{\fih}$ while
$\E_{t'}(\Gamma_1(i'')) \in \bF_{\fih}$  for every $i'' \ne i'$ satisfying $ \dist(i,i'') \le \dist(i,i') $,
we have that $\E_{t'}(i'-1)\neq \E_{t'}(i') $ while $ \E_{t'}(i') = \E_{t'}(i'+1) = \E_{t'}(i'')$ for every
$i''$ as above. This implies that $\E_t(i) = \E_{t'}(i')$ if $\parity(\dist(i',i))=0$ and
$\E_t(i) \neq \E_{t'}(i')$ otherwise.

\item The function $\hrhor$ is defined as follows: $\hrhor(\tau,\beta,\ell) = \tau$ if $\beta=0$ and it equals $\overline{\tau} = \overline{\tau}_1 \overline{\tau}_2 \overline{\tau}_3$, otherwise.

\item
Let $ \sigma : \cycnums{n} \to \bitset $ be a configuration and let $ [x,y] $ be an interval
of locations such that $ \sigma(\Gamma_1(x)) \in  \F_\fih $ and $ \sigma(\Gamma_1(y)) \in \F_\fih $.
We define the configuration $ \tsigma $ as follows.
Let $ i \in \nums(n) $.
If $ i \notin [x,y] $, we set $ \tsigma(i) = \sigma(i) $.
Otherwise, let $ j_i \in [x,y] $ be the location that minimizes $ \dist(i,j) $ among the locations for which $ \sigma(\Gamma_1(j)) \in \F_\fih $, and set $ \tsigma(i) = \sigma(j_i) \oplus \parity(\dist(i,j_i)) $.
In particular, note that if $ \sigma(\Gamma_1(i)) \in \F_\fih $, then $ \tsigma(i) = \sigma(i) $.
We claim that for every $i \in [x,y]$ we have that $ \tsigma(\Gamma_1(i)) \in \F_\fih $, and that if $\sigma(\Gamma_1(i)) \in \F_\fih$, then $ \tsigma(i)=\sigma(i) $.
Let $ i \in [x,y] $.
If $ \sigma(\Gamma_1(i)) \in \F_\fih $, then $ j_i=i $, and hence either $ \sigma(\Gamma_1(i-1)) \in \F_\fih $ or $ \sigma(\Gamma_1(i+1)) \in \F_\fih $, which implies that either $ \tsigma(i) \ne \tsigma(i-1) $ or $ \tsigma(i) \ne \tsigma(i+1) $, and in both cases, it holds that $ \tsigma(\Gamma_1(i)) \in \F_\fih $ and $ \tsigma(i)=\sigma(i) $.
If $ \sigma(\Gamma_1(i)) \in \bF_\fih $, then, assume without loss of generality that $ j^* \in [i+1,y] $ (the case in which $ j^* \in [x,i-1] $ is similar).
In this case, $ j_i = j_{i+1} $, and hence $ \tsigma(i) \ne \tsigma(i+1) $, which implies that $ \tsigma(\Gamma_1(i)) \in \F_\fih $.

\item Assume that $\sigma(\Gamma_1(z)) = 000$ (the case that $\sigma(\Gamma_1(z)) = 111$ is handled analogously).
If $\nu \oplus \gamma' = 0$, then we set $z' = z+1$ and $\tsigma(z'+1) = 1$. Otherwise, we set $z'=z+2$ and $\tsigma(z')=\tsigma(z'+1) = 1$.
The symmetric variant is established analogously.

\end{enumerate}

\subsection{Flip unless homogeneous}

This rule, denoted $\fuh$, is defined as follows: $\fuh(\beta_1,\beta_2,\beta_3) = \overline{\beta}_2$ unless
$\beta_1=\beta_2=\beta_3$ in which case $\fuh(\beta_1,\beta_2,\beta_3) = \beta_2$.
Similarly to $\fuh$, this rule ultimately converges to configurations with blocks of size one or two, where each block flips all values in consecutive time steps (unless the initial configuration is the all-$0$ configuration or the all-$1$ configuration, in which case it remains the initial configuration).
Here too  $\bF_{\fuh} = \{000,111\}$.

\begin{enumerate}
	
	\item  If $\E_t(\Gamma_1(i)) \in \F_{\fuh}$ then either $\E_t(i) \neq \E_t(i-1)$ or $\E_t(i) \neq \E_t(i+1)$ (possibly both), so that in the first case $\E_{t+1}(i-1) \ne \E_t(i-1)$ and in the second case, $\E_{t+1}(i+1) \ne \E_t(i+1)$.
	In both cases $\E_{t+1}(i) \ne \E_t(i)$, so that either $\E_{t+1}(i) \neq \E_{t+1}(i-1)$ or $\E_{t+1}(i) \neq \E_{t+1}(i+1)$.
	That is, $\E_{t+1}(\Gamma_1(i)) \in \F_{\fuh}$ as required.
	
	\item If $\E_t(\Gamma_1(i)) \in \bF_{\fuh}$, then $\E_t(\Gamma_1(i)) \in \{000,111\}$. If both $\E_t(\Gamma_1(i-1))\in \bF_{\fuh}$ and $\E_t(\Gamma_1(i-1))\in \bF_{\fuh}$, then $\E_t(\Gamma_2(i)) \in \{00000,11111\}$, so that $\E_{t+1}(\Gamma_1(i))\in \{000,111\} = \bF_{\fuh}$. On the other hand, if $\E_t(\Gamma_1(i-1))\in \F_{\fuh}$ or $\E_t(\Gamma_1(i+1)) \in \F_{\fuh}$, then $\E_t(\Gamma_2(i)) \in \{1000?,?0001,0111?,?1110\}$ so that $\E_{t+1}\in \{01?,?10,10?,?10\} = \F_{\fuh}$.
	
	\item The function $\frhor$ is defined as follows: $\frhor(\beta_1,\beta_2,\beta_3) = \beta_1 \oplus \beta_2 \oplus \beta_3$.
	This can be verified similarly to what was shown for $ \fih $.
	
	\item The function $\hrhor$ is defined as follows: $\hrhor(\beta,\cdot,\cdot) = \beta$.
	
	\item
	This item is the same as the corresponding one for $ \fih $.
	
	\item This item is very similar to the corresponding one for $ \fih $. The only difference is that there is a dependence on the parameter $ \gamma $, where if $ \gamma = 0 $, the setting is exactly the same as for $ \fih $, and if $ \gamma = 1 $, the setting is the opposite (that is, if $ \nu=0 $, we set $ \tsigma $ the way we did for $ \nu=1 $ in $ \fih $, and vice versa).
	
\end{enumerate}